\documentclass[
11pt,tightenlines,%
nofootinbib,
preprintnumbers,%
amsfonts,%
prd]{revtex4}

\usepackage{graphicx}
\usepackage{color}

\usepackage{amssymb}

\def\0{\over } \def\2{{\textstyle{1\over2}}} \def\4{{\textstyle{1\over4}}}
\def\5{\hat } \def\6{\partial }

\newcommand{\be}{\begin{equation}}
\newcommand{\ee}{\end{equation}}

\newcommand{\bea}{\begin{eqnarray}}
\newcommand{\eea}{\end{eqnarray}}

\newcommand{\nn}{\nonumber\\ }
\newcommand{\g}{g_{\rm eff}}

\newcommand{\geff}{g_{\rm eff}}

\def\Tr{{\,\rm Tr\,}}
\def\Im{{\,\rm Im\,}}
\def\Re{{\,\rm Re\,}}
\def\tr{{\,\rm tr\,}}

\newcommand{\muMS}{\bar\mu_{\overline{\rm MS}}}

\def\cf{C_{f}}

\def\Nf{N_f}

\def\slashchar#1{\setbox0=\hbox{$#1$}           
   \dimen0=\wd0                                 
   \setbox1=\hbox{/} \dimen1=\wd1               
   \ifdim\dimen0>\dimen1                        
      \rlap{\hbox to \dimen0{\hfil/\hfil}}      
      #1                                        
   \else                                        
      \rlap{\hbox to \dimen1{\hfil$#1$\hfil}}   
      /                                         
   \fi}

\newcommand{\nonb}{\nonumber}

\usepackage{bm}
\def\p{{\bf p}}
\def\k{{\bf k}}
\def\q{{\bf q}}

\def\p{{\bf p}}
\def\k{{\bf k}}
\def\q{{\bf q}}

\def\intK{\int_K}
\def\intP{\int_P}
\def\intQ{\int_Q}

\def\intKE{\int_{K_{\rm E}}}
\def\intPE{\int_{P_{\rm_E}}}
\def\intQE{\int_{Q_{\rm_E}}}

\def\intkO{\int_{k_0}}
\def\intpO{\int_{p_0}}

\def\intbfk{\int_{\k}}
\def\intbfp{\int_{\p}}

\def\sumint{\hbox{$\sum$}  \!\!\!\!\!\!\!\int}
\def\sumintp{\hbox{$\sum$} \!\!\!\!\!\!\!\int}

\def\G{G}

\def\K{K}

\def\KM{K}

\def\kOE{k_0}
\def\pOE{p_0}
\def\qOE{q_0}

\def\kOEb{k_0}
\def\pOEb{p_0}

\def\Pil{\Pi_{\ell,\rm th}}

\def\Pivac{\Pi_{\rm vac}}
\def\Pith{\Pi_{\rm th}}

\begin{document}
\preprint{ECT*-05-08}
\preprint{TUW-05-12}
\title{Asymptotic thermal quark masses and the 
entropy of QCD\\
in the large-$N_f$ limit}
\author{Jean-Paul Blaizot}
\affiliation{ECT*, Villa Tambosi, Strada delle Tabarelle 286,\\
I-38050 Villazzano Trento, Italy}
\author{Andreas Ipp}
\affiliation{ECT*, Villa Tambosi, Strada delle Tabarelle 286,\\
I-38050 Villazzano Trento, Italy}
\author{Anton Rebhan}
\affiliation{Institut f\"ur Theoretische Physik, Technische
Universit\"at Wien, \\Wiedner Hauptstr.~8-10, 
A-1040 Vienna, Austria }
\author{Urko Reinosa}
\affiliation{Institut f\"ur Theoretische Physik, Technische
Universit\"at Wien, \\Wiedner Hauptstr.~8-10, 
A-1040 Vienna, Austria }

\begin{abstract}
We study the thermodynamics of QCD in the limit of large flavor
number ($N_f$) and test the proposal to resum the physics of
hard thermal loops (HTL) through a 
nonperturbative expression for the entropy
obtained from a $\Phi$-derivable two-loop approximation.
The fermionic contribution to the entropy involves
a full next-to-leading order evaluation of the asymptotic
thermal quark mass, which is non-local, and for which only
a weighted average value was known previously. For a natural
choice of renormalization scale we find remarkably good agreement of
the next-to-leading-order HTL results for the fermion self energy
and in turn for the entropy with the respective exact large-$N_f$ results
even at very large coupling.
\end{abstract}
\maketitle


\section{Introduction}

The perturbative series for the thermodynamic potentials of hot QCD
is by now known up to and including order $g^6 \log(g)$
\cite{Arnold:1995eb,Zhai:1995ac,Braaten:1996jr,Kajantie:2002wa}, with
$g=\sqrt{4\pi\alpha_s}$ being the Yang-Mills coupling constant.
Taken at face value, this series is poorly convergent and suffers from a strong
dependence on the renormalization point even at temperatures many
orders of magnitude higher than the QCD scale $\Lambda_{\rm QCD}$. 
However, this problem is not specific
to QCD at high temperature, which has
a nonperturbative sector starting to contribute at order $g^6$.
Similarly poor convergence behaviour appears also in
simple scalar field theory \cite{Parwani:1995zz}, and even
in the case of large-$N$ $\phi^4$ theory \cite{Drummond:1997cw},
where all interactions can be resummed in a local thermal mass term,
as soon as one starts
expanding out in a series of powers and logarithms of the coupling.

Various mathematical extrapolation techniques have been tried
to restore the convergence of the perturbative series,
such as  Pad\'e approximants
\cite{Kastening:1997rg,Hatsuda:1997wf,Cvetic:2004xq},
self-similar approximants \cite{Yukalov:2000zr}, and
Borel resummation \cite{Parwani:2000rr,Parwani:2000am}.
A more physically motivated
proposal for reorganizing thermal perturbation theories
is called ``screened perturbation theory'' 
\cite{Karsch:1997gj,Andersen:2000yj}. This is a variant
of a variational perturbation theory, where the tree-level Lagrangian
is modified so that it includes a mass term, which is then
determined by a variational principle (minimal sensitivity).
In the case of gauge theories, the prefactor of the gauge-invariant nonlocal
and nonlinear hard-thermal-loop (HTL)
action \cite{Braaten:1992gm,Frenkel:1992ts} is used
for this purpose in a generalization of this approach 
to QCD by Andersen, Braaten, Petitgirard, and Strickland
\cite{Andersen:1999fw,Andersen:2002ey,Andersen:2003zk,Andersen:2004fp}.

The HTL action is the correct leading-order effective action for soft
modes at energies parametrically smaller than the temperature.
However it is somewhat problematic to also use it at
hard scales, which are responsible for the leading
terms in the thermodynamic potential, albeit the
problems with thermal perturbation theory clearly come from
screening effects at soft scales. Indeed (HTL-)screened perturbation
theory changes the ultraviolet structure and requires
an ad-hoc renormalization of additional UV singularities.

An alternative proposal for resumming the physics of hard thermal
loops was put forward in Refs.\ \cite{Blaizot:1999ip,Blaizot:1999ap,Blaizot:2000fc,Blaizot:2003tw}
\footnote{For an attempt to achieve a comparable
resummation directly in terms of the pressure see \cite{Peshier:2000hx}.}
and is based on
a nonperturbative expression for the entropy density that can be
obtained from a $\Phi$-derivable two-loop approximation 
\cite{Vanderheyden:1998ph}.
Here the emphasis is fully on a quasiparticle picture, whose
residual interactions are assumed to be weak after the
bulk of the interaction effects have been incorporated in
the spectral data of the quasiparticles.
As opposed to methods based on ``screened perturbation theory'', $\Phi$-derivable approximations have better properties with respect to renormalization. They are renormalizable in the case of scalar field theories \cite{vanHees:2001ik,VanHees:2001pf,Blaizot:2003br,Blaizot:2003an,Berges:2004hn,Berges:2005hc}, 
whereas in gauge theories 
the question of renormalizability remains still opened.
However, when combined with HTL approximations, they define UV finite physical quantities without introducing spurious counterterms \cite{Blaizot:1999ip,Blaizot:1999ap,Blaizot:2000fc}.

These approaches indeed succeed in taming the plasmon term $\sim g^3$
that spoils the apparent convergence of strict perturbative expansions
in $g$ and comparison with existing lattice data suggest that
the entropy of QCD for $T \ge 3 T_c$ 
can indeed be accounted for remarkably well when the
leading-order interactions are resummed into
spectral properties of HTL quasiparticles \cite{Blaizot:1999ip,Blaizot:1999ap,Blaizot:2000fc}.

Since this success is 
gauged mostly from the comparison with lattice data,
it is desirable to have other possibilities for testing
these resummation prescriptions. 
In Ref.~\cite{Moore:2002md} 
it was suggested to use the large flavor-number ($N_f$) limit
of QED and QCD for that purpose. The thermodynamic potential
has been worked out in Ref.~\cite{Moore:2002md,Ipp:2003zr,Ipp:2003jy}
to order $N_f^0$ and to all orders in $g^2N_f$, which is
kept finite as the limit $N_f\to\infty$, $g\to 0$ is taken.

Large-$N_f$ QCD is no longer asymptotically free.
In this limit the renormalization scale dependence is determined
(non-perturbatively) by the one-loop beta function according to
\begin{equation}
\label{gscal}
\frac{1}{g_{\rm {eff}}^{2}(\mu )}=\frac{1}{g_{\rm {eff}}^{2}(\mu ')}
+\frac{\log (\mu '/\mu )}{6\pi ^{2}}\,,
\end{equation}
where
we have defined $\g^2 = g^2 N_f/2$ (in QED we would have $\g^2 = g^2 N_f$).
There is a Landau singularity in the
vacuum gauge field propagator at
$ Q^{2}=\Lambda _{\rm L}^{2} $
with
\be
\label{LLandau}
\Lambda_{\rm L}=\bar\mu_{\rm MS}
 e^{5/6}e^{6\pi ^{2}/\g^{2}(\bar\mu_{\rm MS})}\,,
\ee
where $\muMS$ is the scale of modified minimal subtraction.
However, although the large-$N_f$ theory has 
an unavoidable ambiguity associated with its UV completion,
in the thermodynamic potential 
the magnitude of the ambiguity is suppressed
by a factor $(T/\Lambda_{\rm L})^4$, which
in practice means that the Landau pole problem can be ignored
for all coupling $\g(\muMS=\pi T) \lesssim 6$. Since strict perturbation
theory appears to work only for $\g \lesssim 2$, this gives enough
room for testing improvements of thermal perturbation theory.

In this paper we shall test the HTL resummation proposed in 
Refs.\ \cite{Blaizot:1999ip,Blaizot:1999ap,Blaizot:2000fc}
which use the
nonperturbative expression for the entropy
obtained from a $\Phi$-derivable two-loop approximation.
In the large-$N_f$ limit,
a complete HTL resummation involves in particular the
evaluation of next-to-leading-order thermal quark masses
at asymptotic hard scales. In one-loop HTL-resummed perturbation theory,
these thermal masses are nonlocal, i.e.\ functions
of momentum, and so far
only a certain weighted
average was known, and this only to next-to-leading order
in the coupling. In this paper we shall perform a complete
HTL evaluation, which resums an infinite series in the coupling,
and we compare with the numerical evaluation of the exact
large-$N_f$ values. A corresponding evaluation of the
entropy shows considerable improvement compared to
a previous evaluation using the averaged asymptotic
masses at next-to-leading order in HTL-resummed perturbation theory
\cite{Rebhan:2003fj}. Although at larger coupling the HTL results
have a large renormalization scale dependence, 
choosing the scale of fastest apparent
convergence reproduces the exact large-$N_f$ result for all $\g^2$
with a quality comparable to an optimised $g^6$ result \cite{Ipp:2003jy}.

The organization of this paper is as follows:
In section \ref{sec:2PI} we first review the $\Phi$-derivable
approximation to the entropy. At two-loop order we present the
remarkably simple quasiparticle
formulae for the entropy and number density which
can be used both in the large-$N_f$ limit
and for an HTL approximation. In section \ref{sec:largeNf} we
discuss the issue of renormalization both
from the perspective of the thermodynamic potential and
of the quasiparticle entropy and number density. In section
\ref{sec:Sigma} we evaluate the fermion self-energy, which plays
a central role in the quasiparticle entropy expressions,
and which requires to solve a technical problem related to
the necessity of eliminating the Landau pole by a momentum cutoff.
Such a cutoff has to be imposed in a way which respects Euclidean
rotational invariance in order to avoid spurious contributions
in already renormalized expressions [derived in dimensional regularization
as we shall use modified minimal subtraction ($\overline{\rm MS}$)].
Since the quasiparticle entropy is formulated in Minkowski space,
this requirement leads to intricacies in the numerical evaluation,
whose results are finally given
in section \ref{sec:results}. There
we present the numerical results for the complete
large-$N_f$ entropy and the complete momentum-dependent
asymptotic thermal quark masses together with their respective
HTL approximations. We find remarkable agreement of the latter
with the exact results provided the HTL approximation is
used in a nonperturbative manner (i.e., not truncating at
the order of perturbative accuracy)
and an optimized
renormalization scale is chosen. In this case, the agreement
is comparable to optimized results using perturbative results
through order $g^6$.

At this point it is useful to specify the notation to be used
throughout the paper. 
As gauge group we take SU($N$), 
so $N$ denotes the number of quark colors. As already said,
$N_f$ is the number of quark flavors. 
The number of gluons is denoted by $N_g=N^2-1$. 
We use the Minkowski metric $g^{\mu\nu}=(+,-,-,-)$
and denote by capital letters 4-momenta $P$ with components
$P^\mu=(p_0,\p)$, so that $P^2=p_0^2-\p ^2$. In finite
temperature calculations, we are led to set $p_0=i \omega_n$, where
$\omega_n$ is a Matsubara frequency ($\omega_n=n \pi \beta$ where $n$
is an even integer for bosons and odd integer for fermions). We shall
denote the typical sum-integrals that occur at finite temperature by
the following shorthand notation:
\begin{equation}
\sumint_P f(P)=  \frac{1}{\beta}\sum_{n}\,\int\frac{d^3p}{(2\pi)^3} f(i\omega_n,\p)\,.
\end{equation}
At zero temperature, the sum over Matsubara frequencies is replaced by an integral over the imaginary energy axis, leading to an Euclidean integral
denoted by
\begin{equation}
\intPE f(P)  =  \int_{-i \infty}^{i \infty} \frac{dp_0}{2\pi i}\,\int\frac{d^3p}{(2\pi)^3} f(p_0,\p)\,.
\end{equation}
For an integral over the four components of a four-vector,
we simply write
\begin{equation}
\intP  f(P) =  
\int_{-\infty}^{\infty}\frac{dp_0}{2\pi}\,\int\frac{d^3p}{(2\pi)^3} f(p_0,\p) =
\intpO \intbfp f(P)\,.
\end{equation}
We denote the free propagator by
$\Delta_0(P)=-1/(P^2-M^2)$ 
or, in its spectral representation, by
\begin{equation}\label{eq:spectral}
\Delta_0(\omega,p)=\intpO\frac{\rho_0(P)}{p_0-\omega}\,,
\end{equation}
where $\omega$ is a complex energy variable, and the spectral function is $\rho_0(P)=2\pi\epsilon(p_0)\delta(P^2-M^2)$. 
The free fermion propagator is then given by $S_0(P)=(\slashchar{P}-M)\Delta_0(P)$ or, in its spectral form, by
\begin{equation}\label{eq:spectral_fermion}
S_0(\omega,p)=\intpO\frac{(\slashchar{P}-M)\,\rho_0(P)}{p_0-\omega}\, ,
\end{equation}
where we used the fact that $\rho_0(P)$ is an odd function in $p_0$ to replace $\omega\gamma_0$ by $p_0\gamma_0$ in $\slashchar{P}$ within the integral.


\section{$\Phi$-derivable two-loop entropy and HTL resummation}\label{sec:2PI}


\subsection{Generalities}\label{sec:2PI_entropy}

$\Phi$-derivable approximations \cite{Baym:1962} 
are constructed from the two-particle-irreducible (2PI) skeleton expansion
\cite{Luttinger:1960ua,Cornwall:1974vz}.
In the latter, the thermodynamic potential is expressed in
terms of dressed propagators ($\G$ for bosons, $S$ for fermions)
according to
\bea\label{LWQCD}
\Omega[\G,S]&=&\2\, T \Tr \log \G^{-1}-\2\, T \Tr \Pi \G\nn &&
-\,T \Tr \log S^{-1} + T\Tr \Sigma S + T\Phi[\G,S] \, ,
\eea
where ``Tr'' refers to full functional traces 
$\Tr = \tr \int_0^\beta d\tau \int_{\bf x} \rightarrow \beta V \tr  $\raisebox{.4ex}{$\scriptscriptstyle \sum$}$\!\!\!\!\int_K$ 
and
$\Phi[\G,S]$ is the sum of 2-particle-irreducible ``skeleton''
diagrams. In gauge theories one either has to assume a ghost-free
gauge such as temporal axial gauge or
to include the bosonic
Faddeev-Popov ghost propagator $\G_{gh}$.
For simplicity
we shall assume a ghost-free gauge in the general discussion.
At both the level of HTL approximations and in the large-$N_f$
limit, we can and shall use 
the Coulomb gauge for this purpose instead of the
somewhat problematic \cite{James:1990it} temporal axial gauge.

\begin{figure}
\includegraphics[bb=40 310 505 450,scale=0.5]{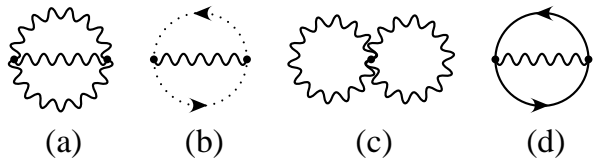}
\caption{Diagrams for $\Phi$ at 2-loop order in QCD. Wiggly, plain, and dotted lines refer respectively to gluons, quarks, and ghosts. Only diagram (d) contributes at next-to-leading order in the large-$N_f$ limit. 
\label{figPhi}}
\end{figure}

Standard contour integration gives
\bea\label{LWQCDci}
\Omega[\G,S]/V&=&\tr \intK n(k_0) \Im \Big[
\log \G^{-1}(\KM)-\Pi(\KM) \G(\KM) \Big] \nn
&&+\,2\tr \intK f(k_0) \Im \Big[
\log  S^{-1}(\KM)-\Sigma(\KM) S(\KM) \Big]+T\Phi[\G,S]/V,\quad
\eea
where ``tr'' denotes a trace over discrete labels only. The Bose-Einstein
and Fermi-Dirac factors are defined as
$n(\omega)=(e^{\omega/T}-1)^{-1}$ and $f(\omega)=(e^{(\omega-\mu)/T}+1)^{-1}$,
respectively.

The self-energies $\Pi=\G^{-1}-\G^{-1}_0$ and 
$\Sigma=S^{-1}-S^{-1}_0$,
where $\G_0$ and $S_0$ are bare propagators, are themselves functionals
of the full propagators. They are determined by the stationarity property:
\be\label{staty}
{\delta \Omega[\G,S]/\delta \G}=0={\delta \Omega[\G,S]/\delta S}\, ,
\ee
according to
\be\label{selfenergies}
{\delta\Phi[\G,S]/\delta \G}=\2\Pi\, ,\quad
{\delta\Phi[\G,S]/\delta S}=-\Sigma\, .
\ee

The $\Phi$-derivable two-loop approximation consists of keeping only
the two-loop skeleton diagrams (see Fig.~\ref{figPhi}), which leads to
a dressed one-loop approximation for the self-energies
(\ref{selfenergies}). In
a gauge theory this generally introduces gauge dependences (which are
parametrically suppressed, though \cite{Arrizabalaga:2002hn}). 
However the further approximations put forward
in Refs.\ \cite{%
Blaizot:1999ip,Blaizot:1999ap,Blaizot:2000fc}
are manifestly gauge independent in the 
$\Phi$-derivable two-loop approximation.

A self-consistent
two-loop approximation for $\Omega$ has a remarkable consequence
for the first derivatives of the thermodynamic potential, the entropy
and the number densities:
\be\label{SNdef}
{\mathcal S}=\left.{\6P\0\6T}\right|_{\mu},\quad
{\mathcal N}=\left.{\6P\0\6\mu}\right|_{T},\quad  P=-\Omega/V.
\ee
Because of the stationarity property (\ref{staty}), one can ignore the $T$ and
$\mu$ dependences implicit in the spectral densities of the full
propagators, and differentiate exclusively the statistical
distribution functions $n$ and $f$ in (\ref{LWQCDci}).
Now the derivative of the {\em two-loop} functional $T\Phi[\G,S]$ at fixed
spectral densities of the propagators $\G$ and $S$ 
turns out to cancel
part of the terms $\Im(\Pi \G)$ and $\Im(\Sigma S)$ in
(\ref{LWQCDci}): 
\bea\label{SP1}
{\cal S}'_{\rm 2-loop}&\equiv& -\left.{\6(T\Phi_{\rm 2-loop})\0\6T}\right|_{\G,S}\nn&&+
\,\tr\intK\left[{\6n(k_0)\0\6T}
\Re\Pi \Im \G + 2{\6f(k_0)\0\6T}
\Re\Sigma\Im S\right]=0 \, ,
\eea
leading to
the remarkably simple 
formulae \cite{Vanderheyden:1998ph,Blaizot:1999ap,Blaizot:2000fc}:
\bea
\label{S2loop}
{\mathcal S}&=&-\tr \intK{\6n(k_0)\0\6T} \Big[ \Im 
\log \G^{-1}-\Im \Pi \Re \G \Big] \nn
&&-2\tr \intK{\6f(k_0)\0\6T} \Big[ \Im
\log  S^{-1}-\Im\Sigma \Re S \Big], \\
\label{N2loop}
{\mathcal N}&=&-2\tr \intK{\6f(k_0)\0\6\mu} \Big[ \Im
\log  S^{-1}-\Im \Sigma \Re S \Big],
\eea
where we have dropped the label ``2-loop'' that could be attached
to 
$\mathcal S$ and $\mathcal N$.

Through these formulae, 
effectively one-loop integrals,
all interactions below perturbative order $g^4$ are
summarized by spectral data only, 
which shows that entropy and density
are the preferred quantities for a quasiparticle description.
In particular the term $g^3$, which usually spoils
the apparent convergence of strict perturbative expansions, is
now incorporated in a nonpolynomial expression, together
with (incompletely resummed) higher-order terms.

Moreover, these expressions are UV finite as soon as 
the self-energies are, and thus the former are useful
as a starting point for further approximations.
In Refs.\ \cite{%
Blaizot:1999ip,Blaizot:1999ap,Blaizot:2000fc} it was proposed
to use the gauge-invariant hard thermal loops for this purpose.
Before considering HTL's, we recall some general properties of self-energies.

\subsection{Bosonic and fermionic self-energies}\label{sec:selfenergy}

In the Coulomb gauge, the gauge propagator can be decomposed into a longitudinal and a transverse contribution
\begin{eqnarray}\label{eq:gluon_prop}
\G_{00}(K) & = & G_{\rm L}(K)\,,\nonumber\\
\G_{ij}(K) & = & \Big\{\delta_{ij}-\frac{k_i k_j}{k^2}\Big\}\,G_{\rm T}(K)\,,
\end{eqnarray}
and we define corresponding self-energy components through
\begin{equation}
G_{\rm L}(K)=\frac{-1}{k^2+\Pi_{\rm L}(K)}\,,\quad
G_{\rm T}(K)=\frac{1}{-K^2+\Pi_{\rm T}(K)}\,.
\end{equation}
Longitudinal and transverse spectral functions are introduced according to
\begin{equation}\label{eq:spectral_gluon}
G_{\rm L}(\omega,k)=-\frac{1}{k^2}+\int_{k_0}\frac{\rho_{\rm  L}(K)}{k_0-\omega}\nonb\,,\quad 
G_{\rm T}(\omega,k)=\int_{k_0}\frac{\rho_{\rm T}(K)}{k_0-\omega}\,.
\end{equation}

In the following it will be convenient to write the temporal component
of the Coulomb gauge
propagator alternatively as
\begin{equation}
\G_{00}(K)=\frac{k_0^2-k^2}{k^2}\G_\ell(K),
\end{equation}
and the
Dyson equations as
\begin{eqnarray}
\G_{\rm T}^{-1} & = & -k_0^2+k^2+\Pi_{t,\rm th}+\Pi_{t,\rm vac}\,,\nonb\\
\G_\ell^{-1} & = & -k_0^2+k^2+\Pi_{\ell,\rm th}+\Pi_{\ell,\rm vac} \, ,
\end{eqnarray}
such that $\Pi_{\rm T}=\Pi_{t,\rm th}+\Pi_{t,\rm vac}$ but
$\Pi_{\ell,\rm th}+\Pi_{\ell,\rm vac} = -\Pi_{\rm L}(k_0^2-k^2)/k^2$.
Here we have separated off the vacuum ($T,\mu\to0$) limit of
the self-energy components.
In the large-$N_f$ limit, we shall have that 
$\Pi_{t,\rm vac}=\Pi_{\ell,\rm vac}\equiv \Pi_{\rm vac}$
even in the non-covariant Coulomb gauge.\\

The most general form of the self-energy $\Sigma$ at finite temperature
and density can be written as 
\begin{equation}
\Sigma(\K)=a(\K)\gamma^{0}+b(\K)\hat{\k}\cdot{\bm\gamma}+c(\K)\, .
\end{equation}
We define the projection of the self-energy on $\slashchar{K}-M$ as
\begin{equation}\label{eq:sigmabar}
\bar{\Sigma}(K)\equiv\tr\left[(\slashchar{K}-M)\Sigma(K)\right]=4\omega a(\omega,k)+4kb(\omega,k)-4 M c(\omega,k)\,.\end{equation}
In the massless case, $c(\K)=0$, and the quark propagator at finite temperature or density can be split into
two separate components with opposite ratio of chirality over helicity:
\begin{equation}
\Delta_\pm(\K)=\frac{1}{-\omega\pm[k + \Sigma_\pm(\K)]}\,,
\end{equation}
where
\begin{equation}\label{eq:sigmapm}
\Sigma_{\pm}(\K)\equiv b(\K)\pm a(\K)\, .
\end{equation}
Furthermore, $\bar{\Sigma}(K)$ and $\Sigma_{\pm}(\K)$
on the light cone $\omega=\pm k$ are related by
\begin{equation}
\Sigma_{\pm}(\omega=\pm k,k)=\frac{1}{4k}\bar{\Sigma}(\omega=\pm k,k)
\end{equation}
(see also Appendix \ref{sec:sigmarelation} for a relation between $\Sigma_+$ and $\Sigma_-$).

\subsection{HTL approximation of the self-consistent entropy}

The HTL effective action \cite{Braaten:1992gm,Frenkel:1992ts}
is an effective action for soft modes with energy scales
$\sim gT$, which are, at least parametrically, at smaller 
energy than the hard modes defined by the temperature scale $T$
(or the chemical potential when $\mu\gg T$).
At hard scales, the HTL effective action is no longer accurate,
except at small virtuality, but this is indeed the
phase space domain which contributes   
the leading-order interaction terms $\propto g^2$ 
in the expressions (\ref{S2loop}) and (\ref{N2loop}). The
order $g^2$ contribution is obtained by expanding out the propagators
and keeping a single self-energy insertion.
In SU($N$) gauge theory with $N_g=N^2-1$ gluons and
$N_f$ massless quarks this leads to
\bea\label{S2}
{\mathcal S}_2 
 &=&2N_g\intK\,{\6n(k_0)\0\6T}\,\Re{\Pi_{\rm T}}
{\Im\frac{1}{k_0^2-k^2}}\nn
&&+4NN_f \intK\,{\6f(k_0)\0\6T}\left[\Re \Sigma_+
\Im{1\0k_0-k} - \Re \Sigma_- \Im{1\0k_0+k}\right]
,
\eea
and similarly in the density expression. 
The imaginary part of the free propagator puts the self-energy insertions
on the light-cone, where they are accurately (to order $g^2$) given
by the HTL value\footnote{Here and in what follows we 
shall denote HTL values by putting
a hat on the quantities in question.} 
of the transverse component of the gluon self-energy $\Pi_{\rm T}$
\be\label{minfty}
\5\Pi_{\rm T}(k_0=k)=\5 m_\infty^2=\2 \5m_D^2\, ,
\ee
and the HTL value of $\Sigma_\pm \equiv 
\2(\vec\gamma\vec k/|\vec k|\pm \gamma_0)\Sigma$,
\be\label{Minfty}
\5\Sigma_\pm(k_0=\pm k)={\5 M_\infty^2\02k}
={\5 M^2\0 k}\, ,
\ee
even though $k$ is no longer soft \cite{Kraemmer:1990dr,Flechsig:1996ju}.
The leading-order terms in the interaction contribution to the
thermodynamic potentials $\mathcal S$ and $\mathcal N$ are thus
related to the ``asymptotic'' thermal masses $m_\infty$
and $M_\infty$ of hard gluons and fermions, respectively.
As indicated in Eqs.~(\ref{minfty}) and (\ref{Minfty}),
at leading order these
are proportional to the HTL Debye mass $\5m_D$ and HTL
fermionic plasma frequency $\5M$, given by
\be
\hat m_D^2=(2N+N_f){g^2 T^2\06}+N_f {g^2\mu^2\02\pi^2},\qquad
\hat M^2={g^2 C_f\08} \left( T^2 + {\mu^2\0\pi^2} \right).
\ee
for (uniform) quark chemical potential $\mu$ and $C_f=(N^2-1)/(2N)$.

However,
higher-order contributions in the thermodynamic potential cannot
be calculated by expanding out the self-energies in the
expressions for the thermodynamic potentials
given in Section \ref{sec:2PI_entropy}
as this would lead to infrared divergences in the electrostatic
sector. At least the Debye mass that appears in the static
propagator needs to be resummed, after which one can calculate
systematically up to and including order $g^6 \log g$.
But already the first contribution from the soft sector,
the ``plasmon term'' $\propto g^3$ spoils the apparent convergence
of a strict perturbative series.

With the above nonperturbative expressions for entropy and density
it is now possible to resum simultaneously the effects of Debye screening
and other collective phenomena, such as dynamical screening
and Landau damping. At soft momentum scales, these are
determined to leading order by the HTL self energy expressions
\bea
\label{PiTL}
\5\Pi_{\rm L}(k_0,k)&=&\5m_D^2\left[1-{k_0\02k}\log{k_0+k\0k_0-k}\right],\nn
\5\Pi_{\rm T}(k_0,k)&=&\frac{1}{2}\left[\5m_D^2+\,\frac{k_0^2 - k^2}{k^2}\,
\5\Pi_{\rm L}\right],
\eea
and
\be\label{SIGHTL}
\hat\Sigma_\pm(k_0,k)\,=\,{\5M^2\0k}\,\left[1\,-\,
\frac{k_0\mp k}{2k}\,\log\,\frac{k_0 + k}{k_0 - k}
\right]\,.\ee

Now Dyson-resumming these self-energies into dressed propagators
and inserting them into the entropy formula (\ref{S2loop}) turns out
to account for only a fraction of the plasmon term $\sim g^3$.
In the entropy formula (\ref{S2loop})
the larger part of the plasmon term arises from {\em hard} momentum
scales, namely from corrections to the leading-order asymptotic masses.

\begin{figure}
\centerline{\includegraphics[bb=145 410 500 465,width=10cm]{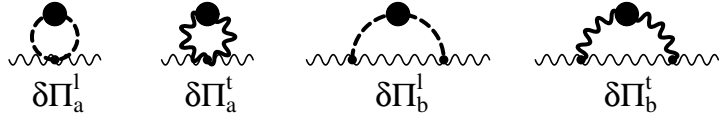}}
\caption{NLO contributions to $\delta\Pi_{\rm T}$ at hard momentum. Thick dashed and wiggly lines with a blob represent HTL-resummed longitudinal and transverse propagators, respectively. In the large-$N_f$ limit the blobs represent full 1-loop resummed propagators. In this limit, these diagrams are suppressed by a factor of $1/N_f$.
\label{fig:dPi}}
\end{figure}

\begin{figure}
\centerline{\includegraphics[bb=15 415 370 465,width=10cm]{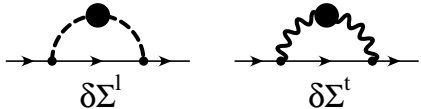}}
\caption{Same as Fig.~\ref{fig:dPi} for the NLO contributions to $\delta\Sigma$ at hard momentum. Although in the large-$N_f$ limit these diagrams are also suppressed individually by a factor of $N_f$, they nevertheless contribute to the entropy at order $N_f^0$.
\label{fig:dSigma}}
\end{figure}

Calculation of the 
next-to-leading-order corrections to the asymptotic thermal masses
(\ref{minfty}) and (\ref{Minfty}) itself requires HTL resummation.
However, because they are given by self-energies with a hard external momentum,
they involve only a single HTL propagator and no HTL vertices,
see Figs.\ \ref{fig:dPi} and \ref{fig:dSigma}.

These corrections to the asymptotic thermal masses are, in contrast
to their leading-order (HTL) values, nontrivial functions of the momentum,
\be\label{dPiSigma}
\delta m_\infty^2(k) \equiv \Re \delta\Pi_{\rm T}(k_0=k),\qquad
\delta M_\infty^2(k) \equiv \Re 
2k\, \delta\Sigma_+(k_0=k).
\ee
At next-to-leading order HTL perturbation theory these expressions can
be evaluated only numerically. However,
through their contribution to the plasmon term, the following
weighted averages are determined to order $g^3$ and given by
remarkably simple results \cite{%
Blaizot:1999ip,Blaizot:1999ap,Blaizot:2000fc}
\bea\label{deltamas}
\bar\delta m_\infty^2 &\equiv&
{\int dk\,k{\6n(k)\0\6T} \delta m_\infty^2(k)
\0 \int dk\,k{\6n(k)\0\6T}}=-{1\02\pi}g^2NT\hat m_D + O(g^4)\;,\\
\label{deltaMas}
\bar\delta M_\infty^2&\equiv&{\int dk\,k{\6f(k)\0\6T} \delta M_\infty^2(k)
\0 \int dk\,k{\6f(k)\0\6T}}=-{1\02\pi}g^2C_fT\hat m_D+ O(g^4)\;.
\eea

Note that these results pertain only to the hard excitations;
corrections to the various thermal masses of soft excitations
are known to differ substantially from (\ref{deltamas}).
For instance, the relative correction to the gluonic
plasma frequency \cite{Schulz:1994gf} at $k=0$,
$\delta m^2_{pl.}/\hat m^2_{pl.}$, 
is only about a third of 
$\bar\delta m_\infty^2/m_\infty^2$; the NLO correction to the
nonabelian Debye mass on the other hand is even positive
for small coupling and moreover logarithmically enhanced
\cite{Rebhan:1993az},
\be
\delta m^2_D = +{1\02\pi}g^2NT\hat m_D \log{c\0g},
\ee
where the constant under the logarithm is nonperturbative and
cannot be calculated by weak-coupling techniques.
However these corrections to the dispersion laws at soft momenta
lead to contributions to the thermodynamic potential which
are beyond the perturbative accuracy of the two-loop
functionals (\ref{S2loop}) and (\ref{N2loop}), and thus we do not
include them in our definition of next-to-leading-order HTL approximation
of the entropy or density. This we 
instead define as an evaluation of (\ref{S2loop}) and (\ref{N2loop})
with HTL propagators and self-energies, which at hard momenta
include the complete corrections to the asymptotic thermal masses
(\ref{dPiSigma}). In Refs.\ \cite{%
Blaizot:1999ip,Blaizot:1999ap,Blaizot:2000fc,Rebhan:2003fj} only the
averaged next-to-leading order asymptotic masses (\ref{deltamas})
and (\ref{deltaMas}) were 
taken into account.

In this paper we shall evaluate
the complete momentum dependence of the next-to-leading-order
asymptotic quark masses 
$\delta M_{\infty}^2$(k), 
which is the only correction also relevant at large $N_f$.
While both (\ref{deltamas})
and (\ref{deltaMas}) are suppressed by $1/N_f$ in the large-$N_f$ limit,
the latter eventually gets
multiplied by a factor $N_f$.
We then compare with the exact
nonperturbative results that one can derive in the 
large-$N_f$ limit 
\cite{Moore:2002md,Ipp:2003zr,Ipp:2003jy}.


\section{Large-$N_f$ expansion of the thermodynamic potential}\label{sec:largeNf}

We shall now focus on the large-$N_f$ limit of the thermodynamic
potential, which is nonperturbative in the effective coupling constant
and therefore requires nonperturbative renormalization.
In this respect we extend the previous
derivations of Ref.~\cite{Moore:2002md,Ipp:2003zr,Ipp:2003jy}
to include nonzero quark masses (see also \cite{Aarts:2005vc}),
even though the final numerical evaluation will be carried out
only for the massless case.
We then discuss the special features of the self-consistent entropy
and density expressions in the large-$N_f$ limit.


\subsection{Large-$N_f$ limit of the 2PI thermodynamic potential}\label{sec:largeNf2PI}

The thermodynamic potential of Eq.~(\ref{LWQCD}) is given in terms of full
propagators $\G$ and $S$ and the $\Phi$ functional, the sum of all 2PI diagrams. 
In the large-$N_f$ limit, the $\Phi$ functional reduces to
a single two-loop skeleton diagram, namely the last one of those displayed in
Fig.~\ref{figPhi}. Only this diagram remains of order 1 when
$g^2\to 0$, $N_f\to \infty$ with fixed $\g^2 \propto g^2 N_f \sim O(1)$.
Furthermore, the fermion propagator in this diagram can be
replaced by the free one, since a fermion self energy is
of order $\Sigma \sim 1/N_f$. Hence
\be
\Phi[\G,S]\to \Phi_*[\G] = \2 \Tr(\Pi_* \G)
\ee
where $\Pi_*$ is a fermion bubble involving the
free fermion propagator $S_0$, and is proportional to $\g^2$. From a variational
point of view, $\Pi_*$ is now a fixed quantity, and the
self-consistent $\Pi$ defined through Eq.~(\ref{selfenergies})
is equal to it because $\Phi_*[\G]$ is linear in $\G$. Hence,
$\Pi=2 \delta \Phi_*[\G]/\delta \G=\Pi_*$. On the other hand,
$S = S_0$ is no longer a variational quantity at all. We thus have
\be\label{eq:before_eval}
\Omega[\G,S]\to \Omega_*[\G]=- T \Tr \log S^{-1}_0+\2 T \Tr \log \G^{-1}
-\2 T \Tr(\Pi\G)+\Phi_*[\G]\, ,
\ee
which, when evaluated at its extremal point and setting $\G_{*}^{-1}=G_0^{-1}+\Pi_*$, 
reduces to\footnote{We stress here that viewed as a functional of $G_*\rightarrow G$, 
the right hand side of Eq.~(\ref{eq:eval_extr}) is not stationary at 
$G_*^{-1}=G_0^{-1}+\Pi_*$ and thus one cannot use the same simplifications to 
compute the entropy as in the 2PI framework. In order to do so, one should rather 
work with Eq.~(\ref{eq:before_eval})}
\be\label{eq:eval_extr}
\Omega_*[\G_*]=-PV=- T \Tr \log S^{-1}_0+\2 T \Tr \log \G_{*}^{-1}.
\ee

\begin{figure}
\begin{center}
\includegraphics[width=7cm]{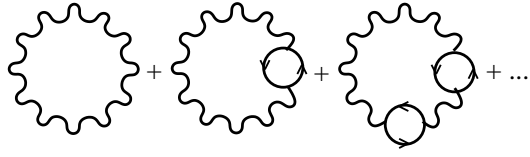}
\caption{
Fermion bubble resummation contributing to the pressure to next-to-leading 
order in the $1/\Nf$ expansion.}\label{fig:bubble}
\end{center}
\end{figure} 

Diagrammatically,
the pressure at leading order ($N_f^1$)
is thus just given by the free fermion loop. 
It contains an overall, temperature independent divergence which can be 
eliminated by subtracting the corresponding vacuum pressure:
\begin{equation}
P_0^f = N\Nf \, \tr \left[ \sumintp_{K} \; \, \log S_0^{-1} \; 
        - \intKE \, \log S_0^{-1} \right]\,.
\label{eq:PLO}
\end{equation}
In Eq.~(\ref{eq:PLO}) and in the rest of this section (\ref{sec:largeNf2PI}), we perform explicitly flavor 
and color traces. Thus the symbol ``$\tr$'' only denotes a trace over spin indices. 
The next-to-leading order contribution $\sim N_f^0$ is obtained 
from the resummation of
the bosonic self-energy $\Pi$ (see Fig.~\ref{fig:bubble}) 
on the gluon propagator (ring resummation):
\begin{equation}\label{eq:bubble}
\Pi_{\rm b}^{\mu\nu}(Q)=\g^2 \, \tr \sumintp_{K} \; \, \gamma^\mu\,S_0(K)\,
\gamma^\nu\,S_0(K-Q)\,,
\end{equation}
which one can split into a vacuum and a thermal piece 
(see Appendix \ref{sec:Matsubara_gluon}):
\begin{eqnarray}\label{eq:decomp}
\Pi_{\rm b,\,vac}^{\mu\nu}(Q) & = & \g^2\, \tr\, \intKE \; \, 
\gamma^\mu\,S_0(K)\,\gamma^\nu\,S_0(K-Q)\,,\nonb\\
\Pith^{\mu\nu}(Q) & = & 2\,\g^2 \, \tr \intK \; \, 
\gamma^\mu\,\left(\slashchar{K}-M\right)\,\sigma_0(K)\,\gamma^\nu\,S_0(K-Q)\,,
\end{eqnarray}
with $\sigma_0(K)=\epsilon(k_0)\,f(|k_0|)\,\rho_0(K)$.
 
This function $\sigma_0(K)$ is a particular combination of the thermal factor and 
the spectral density which appears systematically when isolating thermal 
contributions in the Matsubara formalism \cite{Blaizot:2004bg}. The thermal factor 
$f(|k_0|)$ only involves positive energies and thus the thermal contribution 
is UV finite. We will also need the dominant UV asymptotic behaviour of 
$\Pith^{\mu\nu}$. Since the momentum $K$ is cut off by the temperature, 
the leading asymptotic behaviour at large $Q$ is 
a priori determined by $S_0(K-Q)\sim 1/Q$. 
However, since the gluon self-energy 
is an even function of $Q$, this asymptotic behaviour is in fact improved to 
$1/Q^2$~.\footnote{A more detailed analysis of the asymptotic behaviour can 
be found in the appendix of Ref.~\cite{Moore:2002md}.}

The vacuum piece $\Pi_{\rm b,\,vac}^{\mu\nu}$
is UV divergent (hence the label ``b'' standing for ``bare''). 
However, 
to the same order in $\Nf$, one can 
add the counterterm $\delta Z_A \{Q^2\,g^{\mu\nu}-Q^\mu Q^\nu\}$,
with $\delta Z_A$ adjusted in order 
to absorb the divergence. 
The UV finite self-energy then reads:
\begin{equation}\label{eq:Pivac_ren}
\Pivac^{\mu\nu}(Q)=\g^2 \, \tr \intKE \; \, \gamma^\mu\,S_0(K)\,
\gamma^\nu\,S_0(K-Q)+\delta Z_A \Big\{Q^2\,g^{\mu\nu}-Q^\mu Q^\nu\Big\}\,.
\end{equation}

\begin{figure}
\begin{center}
\includegraphics[width=5cm]{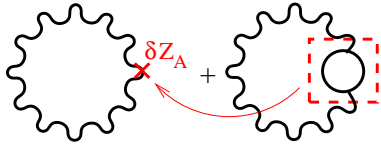}
\caption{Role of the counterterm $\delta Z_A$: every time that a fermion 
bubble is inserted, one has to insert the corresponding counterterm 
$\delta Z_A$ to absorb the subdivergence.\label{fig:div1}}
\end{center}
\end{figure} 
\begin{figure}
\begin{center}
\includegraphics[width=1.5cm]{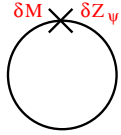}
\caption{Fermion ``counterterms'' eliminating next-to-leading 
order divergences in the bubble resummation of Fig.~\ref{fig:bubble}.
\label{fig:bubble2}}
\end{center}
\end{figure} 
\begin{figure}
\begin{center}
\includegraphics[width=4cm]{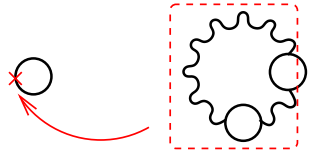}
\caption{Potential singularity with the structure of a fermion self-energy 
insertion. It appears when one of the fermionic momenta is kept fixed while 
all others momenta are taken to infinity. The structure of the diagram 
needed to absorb this potential singularity is exactly that of the diagram 
in Fig.~\ref{fig:bubble2}.\label{fig:div2}}
\end{center}
\end{figure} 
The role of $\delta Z_A$ in the ring resummation is illustrated in Fig.~\ref{fig:div1}.
In general, the ring resummation does not
account for all the next-to-leading order contributions.
Indeed, the fermionic mass and field strength counterterms 
($\delta M$ and $\delta Z_\psi$) are of order $1/\Nf$, 
and can be inserted in a fermion loop (see Fig.~\ref{fig:bubble2}) 
to generate a contribution of order $\Nf^0$. This contribution is 
needed to remove potential subdivergences as 
illustrated in Fig.~\ref{fig:div2}.

Once all subdivergences are eliminated, there remains an overall, 
temperature independent divergence in the pressure, which can be 
eliminated by subtracting the vacuum pressure at next-to-leading order:
\begin{eqnarray}
P-P_0^f & = & N\Nf\tr \left[\sumintp_{K}\;
\left(\delta M+\slashchar{K}\delta Z_{\psi}\right)\,S_0
-\intKE\;\left(\delta M+\slashchar{K}\delta Z_{\psi}\right)\,S_0\right]\nonb\\
& & -\,\frac{N_g}{2} \tr \left[ \sumint_{Q}
        \; \log \left( \G^{-1}_0 
        + \Pi \right) \; \;
        - \intQE \, \log \left( \G^{-1}_0 
        + \Pi_{\rm vac} \right)
        \right] \,.
\label{eq:PNLO}
\end{eqnarray}
It is perhaps surprising that the first line of this formula did not 
appear in the general derivation given at the beginning of this section. 
This is because, the derivation there was made using bare quantities. If 
one rewrites Eq.~(\ref{eq:eval_extr}) in terms of renormalized quantities, 
one generates these extra terms. Performing the Matsubara sums in 
Eq.~(\ref{eq:PNLO}) one obtains:
\begin{eqnarray}
P-P_0^f & = & N\Nf\tr \int_{K}\;\left(\delta M+\slashchar{K}
\delta Z_{\psi}\right)\,(\slashchar{K}-M)\,\sigma_0(K)
-N_g\tr \int_{Q}\epsilon(q_0)n(|q_0|)
\Im \log\left(\G^{-1}_0+\Pi\right)\nonb\\
& - & \frac{N_g}{2} \tr  \intQE\Big[
        \; \log \left( \G^{-1}_0 
        + \Pi \right) \; \;
        -  \, \log \left( \G^{-1}_0 
        + \Pi_{\rm vac} \right) \Big]
        \,.
\label{eq:PNLO2}
\end{eqnarray}

The pressure in Eq.~(\ref{eq:PNLO2}) contains a contribution 
carrying the counterterms, a second contribution which is explicitly 
finite due to the presence of the factor $n(|q_0|)$, and a potentially 
divergent part which reads ($A \simeq B$ here means that the divergent parts of $A$ and $B$ coincide):
\begin{equation}
(P-P_0^f)^{\rm div}\simeq -\frac{N_g}{2} \tr \intQE
        \;  \Big[\log \left( \G^{-1}_0 
        + \Pi \right) \; \;
        - \, \log \left( \G^{-1}_0
        + \Pi_{\rm vac} \right)
        \Big]\,,
\end{equation}
or, after performing the traces explicitly (see also \cite{Ipp:2003zr}), 
\begin{equation}
(P-P_0^f)^{\rm div} \simeq -N_g \intQE \; 
\left[\log \left(1+\frac{\Pi_{t,\rm th}}{-Q^2+\Pi_{\rm vac}}\right)
-\frac{1}{2} \log \left(1+\frac{\Pil}{-Q^2+\Pi_{\rm vac}} \right)\right]\,.
\end{equation}
Using the asymptotic behaviour of $\Pi_{t,\rm th}$ and $\Pil$ which 
can be found in the Appendix of Ref.~\cite{Moore:2002md},
one can show that the potential divergent term arises from the leading terms 
in the expansion of the logarithms:
\begin{equation}\label{eq:bob}
(P-P_0^f)^{\rm div} \simeq -\frac{N_g}{2}\intQE \; 
\frac{\Pil+2\Pi_{t,\rm th}}{-Q^2+\Pi_{\rm vac}}=
-\frac{N_g}{2}\intQE \; \Pi_{\rm th}^{\mu\nu}\G^{\rm vac}_{\mu\nu}\,,
\end{equation}
or using the expression (\ref{eq:decomp}) for 
$\Pi^{\mu\nu}_{\rm th}$ and $C_f=N_g/(2N)$:
\begin{equation}
(P-P_0^f)^{\rm div} \simeq N\Nf\tr\intK \;(\slashchar{K}-M)\,\sigma_0(K)\,
\left[-g^2\cf\intQE \; \gamma^\nu\,S_0(K-Q)\gamma^\mu\,\G^{\rm vac}_{\mu\nu}(Q)\right]\,.
\end{equation}
As we shall see in Sect.~\ref{sec:fermion_vac},
in the process of performing the analytic continuation in $k_0$
of the quantity in brackets in the formula above, one has to deform the 
contour of the $Q_E$ integration in order to avoid crossing singularities. 
However, here we are only interested in the UV 
contributions. Since
contour deformation will only be necessary for 
$Q_E\leq 2k=2k_0$
as we will see at the end of section \ref{sec:cutoffexample}, we can 
identify the integral over $Q_E$ with the vacuum fermion self-energy (see Eq. (\ref{eq:sigmavac}) below), and write:
\begin{equation}
(P-P_0^f)^{\rm div} \simeq N\Nf\tr\intK \;\left(\slashchar{K}-M\right)\,\sigma_0(K)\,
\Sigma_{\rm b,\,vac}[G_{\rm vac}](K)\,\,.
\end{equation}
We now add this divergent contribution to  
that of the counterterms $\delta M$ and $\delta Z_\psi$ and get
\begin{equation}\label{eq:divct}
(P-P_0^f)^{\rm ct + div} \simeq N\Nf\tr\intK \;\left(\slashchar{K}-M\right)\,
\sigma_0(K)\,\Big[\delta M+\slashchar{K}\delta Z_\psi+\Sigma_{\rm b,\,vac}[G_{\rm vac}](K)\Big]\,,
\end{equation}
which can be made finite by a suitable adjustment of $\delta M$ and $\delta Z_\psi$.

In the massless case 
the divergences associated to the fermionic self-energies do not 
contribute to the pressure. 
As $M$ vanishes, and $\slashchar{K}=0$, it is clear that $\delta M$ does not contribute.
One can verify that $\delta Z_\psi$ does not contribute in two ways. From (\ref{eq:divct}) we have:
\begin{equation}\label{eq:PNLOct}
(P-P_0^f)^{\rm ct}\longrightarrow N\Nf\,\delta Z_\psi\,\tr\intK \;\slashchar{K}\,\sigma_0(K)\,\slashchar{K}=2\pi\,N\Nf\,\delta Z_\psi\,\tr\intK \;f(|k_0|)\,\delta(K^2)\,K^2=0\,.
\end{equation}
Equivalently, one can rewrite the potentially divergent piece as:
\begin{equation}\label{eq:PNLOdiv}
(P-P_0^f)^{\rm div}\longrightarrow N\Nf\,\tr\intK \;\slashchar{K}\,\sigma_0(K)\,\Sigma_{\rm b,\,vac}[G_{\rm vac}](K)=2\pi\,N\Nf\,\intK \;f(|k_0|)\,\delta(K^2)\,\bar\Sigma_{\rm vac}[G_{\rm vac}](K)\,.
\end{equation}
As we shall explain in section \ref{sec:ren}, 
$\bar\Sigma_{\rm vac}[G_{\rm vac}]\equiv 
\tr (\slashchar{K} \Sigma_{\rm vac}[G_{\rm vac}])$ 
vanishes on the light cone. This result is in one-to-one correspondence with 
the fact that, in the massless case, the large-$Q$ contribution of the integrand 
in (\ref{eq:bob}), vanishes after integration over $Q$ 
\cite{Moore:2002md,Ipp:2003zr}. This is due to the vanishing of the angular integral 
and requires that numerical computations use 
a cutoff respecting Euclidean symmetry.
Although we shall eventually work with renormalized quantities
obtained in the symmetry-preserving dimensional regularization,
the numerical evaluations require a cutoff to eliminate the
Landau pole of the large-$N_f$ theory. As we shall discuss
further in Sect.~\ref{sec:Sigma}, if this cutoff 
does not respect Euclidean rotation invariance, 
it leads to spurious contributions.
 
With renormalized self energy $\Pi$, Eq.~(\ref{eq:PNLO2}) reduces in the massless case to
\cite{Moore:2002md,Ipp:2003zr,Ipp:2003jy}:
\be
P-P_0^f  =   -  \frac{N_g}{2} \tr \left[ \sumint_{Q}
        \; \log \left( \G^{-1}_0 
        + \Pi \right) \; \;
        - \intQE \, \log \left( \G^{-1}_0 
        + \Pi_{\rm vac} \right)
        \right] \,.
\label{eq:PNLOml}
\ee

\subsection{Large-$N_f$ limit of the self-consistent entropy}\label{sec:2PI_Nf}

The entropy could be obtained by
taking a total derivative of the pressure with respect to the temperature.
Using Eq.~(\ref{eq:PNLOml}) for the pressure, this leads to
\be\label{SlnfO}
\mathcal S-\mathcal S_0 =
{d(P-P_0)\0dT}=-\tr \intK{\6n(k_0)\0\6T}
\Im \log (\G^{-1}\,\G_0)-\tr \intK n(k_0)
\Im \left(\G{d\Pi_*\0dT}\right)
\ee
where $\G^{-1}=\G_0^{-1}+\Pi_*$
and we subtracted the fermionic and the bosonic interaction-free contributions 
to the pressure $P_0 = P_0^f + P_0^b$. 
Note that the overall temperature-independent divergence that is
present in the pressure disappears when considering the entropy. 
As we shall see, the only remaining divergences are associated with
the self energies. This is, however, not manifest in Eq. (\ref{SlnfO}),
where the second term is potentially divergent for $k_0\to-\infty$.

Using that $\Phi_*[\G]$ is a two-loop diagram for which $\mathcal S'$
as defined in Eq.~(\ref{SP1}) vanishes identically, we have
\be\label{SP1lnf}
\left.{\6(T\Phi_*[\G])\0\6T}\right|_{\G}=\tr
\intK\left[{\6n(k_0)\0\6T}
\Re\Pi_* \Im \G + 2{\6f(k_0)\0\6T}
\Re\Sigma\Im S_0\right]\, ,
\ee
where $\Sigma$ is the $O(N_f^{-1})$ fermionic self-energy obtained by
calculating the one-loop diagram composed of a bare fermion propagator
and a full large-$N_f$ gauge boson propagator (see section \ref{sec:Sigma}).

On the other hand,
\bea\label{dPhilnf}
\left.{\6(T\Phi_*[\G])\0\6T}\right|_{\G}&=&{\6\0\6T}\left.\left({T\02}\Tr(\Pi_* \G)\right)
\right|_{\G}={\6\0\6T}\left.\left(\tr \intK n(k_0)
\Im (\Pi_*G)\right)\right|_{\G}\nn
&=&\tr \intK{\6n(k_0)\0\6T}
\Im (\Pi_* \G)+\tr \intK n(k_0)
\Im \left(\G{d\Pi_*\0dT}\right)\, .
\eea
Combining Eqs.~(\ref{SlnfO}), (\ref{SP1lnf}), and (\ref{dPhilnf})
yields
\bea
\label{Slnf}
{\mathcal S}-{\mathcal S}_0&=&-\tr \intK{\6n(k_0)\0\6T}\, \Big[ \Im 
\log (\G^{-1}G_0)-\Im \Pi \Re \G \Big] \nn
&&-2\tr \intK{\6f(k_0)\0\6T}\,
\Re \Sigma \Im S_0\,,
\eea
where $\mathcal S_0$ denotes the ideal-gas limit of the entropy.

This expression can in fact be derived also from the
2-loop-$\Phi$-derivable expression for the entropy, Eq.~(\ref{S2loop}).
In the large-$N_f$ limit, 
the full gauge boson self-energy $\Pi$
reduces to the fermion loop involving undressed
propagators. Any fermion self energy diagram insertion 
(Fig.~\ref{fig:dSigma}) in $\Pi$ would
bring in a factor $g^2$ without a factor $N_f$,
leading to a subleading correction.
However, one fermion self-energy
has to be included in $\tr\log S^{-1} = \tr\log(S_0^{-1}+\Sigma)$
in order to produce contributions of order $N_f^1$ and $N_f^0$.
The integrand of the fermionic terms in
the 2-loop entropy and density, Eqs.~(\ref{S2loop}) and (\ref{N2loop}),
therefore simplifies according to
\bea
\Im
\log  S^{-1}-\Im ( \Sigma) \Re (S )
& \to &
\Im \log S_0^{-1} + \Im (\Sigma S_0)  -  \Im \Sigma \Re S_0 + O(N_f^{-1})\nn
& \to &
\Im
\log  S_0^{-1}+\Re \Sigma \Im S_0\,.
\eea
This enables us to recover
Eq.~(\ref{Slnf}) for the entropy, while we get
\be
\label{Nlnf}
{\mathcal N}-{\mathcal N}_0=-2\tr \intK{\6f(k_0)\0\6\mu}\, 
\Re \Sigma \Im S_0
\ee
for the contribution of order $N_f^0$ to the density. 

In contrast to Eq.~(\ref{SlnfO}),
Eq.~(\ref{Slnf}) is a manifestly ultraviolet finite expression as soon as
propagators and self-energies are made finite through renormalization,
and it naturally identifies 
bosonic and fermionic quasiparticle entropies, $\mathcal S=\mathcal S_b
+\mathcal S_f$, with vanishing residual interactions $\mathcal S'$.


\subsection{HTL approximation of the large-$\Nf$ entropy}

In the large-$\Nf$ limit, the two expressions for the entropy,
Eqs.~(\ref{SlnfO}) and Eq.~(\ref{Slnf}), are completely equivalent.
However, this equivalence is broken once one starts
approximating the gluon self-energy $\Pi_*$ by its HTL limit
which has different behaviour at large momentum.
Doing so leads to uncancelled UV divergences in Eq.~(\ref{SlnfO}),
whereas the entropy formula (\ref{Slnf}) that was derived
using the stationarity of the thermodynamic potential for
self-consistent solutions remains finite. In the following
we shall investigate the quantitative difference it makes to
approximate the full large-$N_f$ gluon self-energy by its
HTL approximation and define
\bea
\label{SlnfHTL}
\hat {\mathcal S} &=&{\mathcal S}_0-\tr \intK{\6n(k_0)\0\6T} \left[ \Im 
\log (\hat G^{-1}G_0)-\Im \hat \Pi  \Re \hat G \right] \nn
&&\hspace{0.5cm}-\,2\tr \intK{\6f(k_0)\0\6T}\, 
\Re \Sigma[\hat G] \Im S_0=
{\mathcal S}_0+\hat {\mathcal S}_b+\hat {\mathcal S}_f\,,
\eea
where $\hat\Pi$ and $\hat G$ denote the gluon HTL self-energy and propagator,
respectively.
The numerical evaluation of this expression does not contain 
spurious divergences, because $\Re \Sigma[\hat G]$ on the light cone 
is UV finite as we will show in section \ref{sec:ren}. This is the 
advantage of using the self-consistent 2PI formulation rather than 
a direct ring resummation. 

It is in fact
instructive to evaluate the difference between Eq.~(\ref{SlnfHTL})
and a direct HTL approximation
of 
$d(P-P_0)/dT$
as given by Eq.~(\ref{SlnfO}). 
In the large-$\Nf$ limit the equivalence of the two entropy formulae 
stems from Eq.~(\ref{SP1lnf}) and Eq.~(\ref{dPhilnf}) which combine together into:
\be
\tr \intK{\6n(k_0)\0\6T}\Im \Pi_* \Re \G-2\tr\intK{\6f(k_0)\0\6T}\Re\Sigma\Im S_0
=-\tr \intK n(k_0)\Im \left(\G{d\Pi_*\0dT}\right)\, .
\ee
In the HTL approximation, it is possible to derive a similar formula 
by exploiting the fact that Eq.~(\ref{SP1lnf}) and Eq.~(\ref{dPhilnf}) 
are valid for any $G$ provided one 
maintains the functional relation of $\Sigma$ to $G$.
The value of $\Pi_*$ does not need to be changed. Thus, if we replace 
$G\to\hat G$, $\Sigma\to\Sigma[\hat G]$, we obtain
\be
\tr \intK{\6n(k_0)\0\6T}\Im \Pi_* \Re \hat G-2\tr\intK{\6f(k_0)\0\6T}
\Re\Sigma[\hat G]\Im S_0=-\tr \intK n(k_0)\Im \left(\hat G{d\Pi_*\0dT}\right)\, .
\ee
Now we add a common term to both sides in order to obtain the 2PI HTL 
entropy in the form:
\bea
\hat \mathcal S & = & {d\0dT}\left(
(P-P_0^f)|_{\Pi_*\to\hat\Pi}\right)
+\tr \intK{\6n(k_0)\0\6T}\Im (\hat\Pi-\Pi_*) \Re \hat G\nonumber\\
&  & \hspace{0.95cm}+\,\tr \intK n(k_0)\Im \left[\hat G\left({d\hat\Pi\0dT}-{d\Pi_{\rm *}\0dT}\right)\right]\,.
\eea
The last two terms are the ones corresponding to the temperature 
dependent singularities in a direct HTL approximation on ring diagrams. 

Returning to the expression (\ref{SlnfHTL}), we observe
that the large-$N_f$ limit leads to rather different simplifications
in the bosonic and the fermionic contributions. In the bosonic
part $\hat{\mathcal S}_b$,
the NLO contributions $\delta\Pi$ displayed in Fig.~\ref{fig:dPi}
do not appear because they are of order $N_f^{-1}$. The NLO
contribution $\delta\Sigma$ of Fig.~\ref{fig:dSigma} is also
of order $N_f^{-1}$, but it does contribute
to the fermionic entropy because there are $N_f$ fermions.
However, $\delta\Sigma$ is not Dyson-resummed because for
$N_f\to\infty$ only one insertion of $\delta\Sigma$ survives.

Because of the factor $\Im S_0$ the fermion self-energy in (\ref{Slnf})
and (\ref{SlnfHTL})
is evaluated
on mass shell and because the integral is dominated by
hard momenta, the fermionic contribution to the entropy 
can be identified with a weighted average over a
momentum-dependent asymptotic quark mass.
For the entropy\footnote{For the quark density a different weighted
average would be relevant.} 
at zero chemical potential, we can define
\bea\label{Minftybarfull}
({\mathcal S}-\mathcal S_0)_{f}
&=&-2\tr \intK{\6f(k_0)\0\6T}
\Re \Sigma \Im S_0\nn
&=&-4NN_f \intbfk{\6f(k)\0\6T}{M_\infty^2(k)\02k}
\equiv -{NN_f T\06} \bar M_\infty^2\,,
\eea
although, as we shall see, 
$M_\infty^2(k)$ and $\bar M_\infty^2$ do not need to be
positive. 

In perturbation theory we have in the massless case
and at zero chemical potential
\be\label{Minftypth}
N_f \bar M_\infty^2 = \left[
{\g^2\02} - {\g^3\0\sqrt3\pi} + O(\g^4)\right]C_f T^2,\quad 
C_f=\frac{N_g}{2N}
\ee
where the calculation
of the contribution $\propto g^3$ requires HTL resummation
as shown in Fig.~\ref{fig:dSigma}.
The latter is responsible for 3/4 of the plasmon term in
the thermodynamic potential;
the remaining 1/4 of the plasmon term comes from the soft
momentum regime of the bosonic entropy (\ref{Slnf}).

In the following we shall 
derive and evaluate numerically the quantities
$M_\infty^2(k)$ and $\bar M_\infty^2$ in the large-$N_f$ limit.
These will be compared with strictly perturbative results and
HTL approximations where all the higher-order terms generated
by HTL resummation are retained.
This will allow us to test the proposal of complete HTL resummation
not only with respect to the entropy of large-$N_f$ QCD but also
with respect to the asymptotic thermal quark masses.


\section{Fermion self-energy}\label{sec:Sigma}

In this section we shall calculate the fermionic self-energy $\Sigma$ in the large-$N_f$ limit as it is needed in the self-consistent entropy and number density formulae Eqs.~(\ref{Slnf}) and (\ref{Nlnf}).
It corresponds to the one-loop diagram of Fig.~\ref{fig:dSigma} with an undressed fermion line and a dressed gluon propagator which resums the fermion bubbles in analogy to 
Fig.~\ref{fig:bubble}. The fermion self-energy 
requires mass and wave function renormalization: $\Sigma(K)=\Sigma_{\rm b}(K)+\delta M+\slashchar{K}\delta Z_\psi$ with 
the bare self-energy $\Sigma_{\rm b}$ given by
\begin{equation}\label{eq:fermion}
\Sigma_{\rm b}(K)=-g^2\cf\sumint_{Q} \; \, \gamma^\mu\,S_0(K-Q)\,\gamma^\nu\,\G_{\mu\nu}(Q)\,.
\end{equation}
It enters the fermionic contribution to the entropy in the second line of Eq.~(\ref{Slnf}) as
\begin{equation}
\mathcal{S}^f-\mathcal{S}_0^f=-N\Nf\,\intbfk\,\frac{1}{2\varepsilon_k}\left[\frac{\partial f(\varepsilon_k)}{\partial T}\,\Re \bar\Sigma(\varepsilon_k,k)-\frac{\partial f(-\varepsilon_k)}{\partial T}\,\Re \bar\Sigma(-\varepsilon_k,k)\right]\,,
\end{equation}
with $\bar\Sigma(K)$ from Eq.~(\ref{eq:sigmabar}).
In the massless case this reduces to (see Appendix \ref{sec:sigmarelation}): 
\begin{equation}
\mathcal{S}^f-\mathcal{S}_0^f=-N\Nf\,\intbfk\,\frac{1}{k}\frac{\partial f(k)}{\partial T}\,\Re \bar\Sigma(k_0 = k,k)\,.
\end{equation}

We note here that the frequency carried by $K$ in Eq.~(\ref{eq:fermion}) is imaginary (Matsubara frequency). In contrast the evaluation of the entropy needs the discontinuity of the fermion self-energy across the real axis. The analytical continuation of (\ref{eq:fermion}) is described in detail in section \ref{sec:fermion_vac}. To that purpose, it is convenient to split the self-energy into a vacuum and a thermal piece $\Sigma_{\rm b}=\Sigma_{\rm b,\,vac}+\Sigma_{\rm th}$  
with respect to the overall loop momentum $Q$. 
The thermal piece is finite 
and its analytical continuation causes no troubles (see section \ref{sec:fermion_Matsubara}). In contrast, the correct evaluation of the vacuum part
with respect to the integral over $Q$ needs that we maintain Euclidean invariance (see section \ref{sec:fermion_vac}). We thus compute it by
replacing the discrete sum in (\ref{eq:fermion}) 
by a continuous integral, leading to a 4-dimensional Euclidean integral:
\begin{equation}\label{eq:sigmavac}
\Sigma_{\rm b,\,vac}(K)=-g^2\cf\intQE \; \, \gamma^\mu\,S_0(K-Q)\,\gamma^\nu\,\G_{\mu\nu}(Q)\,.
\end{equation}
This contribution also carries temperature dependences, but only
through the gluon propagator. It contains UV divergences which in the massless case are washed out from the entropy formula as it is clear by looking at the contribution of the counterterms (once again the trace is performed over the Lorentz group):
\begin{eqnarray}\label{eq:sigmact}
\Re \bar \Sigma^{\rm ct}(\pm \varepsilon_k,k) & = & \Re\tr\Big[\left(\slashchar{ K}-M\right)\,(\delta M+\slashchar{K}\delta Z_\psi)\Big]_{k_0=\pm\varepsilon_k}\nonb\\
& = & \pm4M^2\delta Z_\psi-4M\delta M\,.
\end{eqnarray}
In section \ref{sec:ren}, we give a direct proof of the finiteness of $\Re \bar \Sigma_{\rm b,\, vac}$. In what follows, we drop the label ``b'' on the projected self-energy which reads:
\begin{equation}\label{eq:sigma}
\bar{\Sigma}(K)= \tr \left[\slashchar{K}\Sigma(K)\right]
=-g^2C_f\,\sumint_{Q}\;\tr\Big[\slashchar{K}\,\gamma^\mu\,\slashchar{P}\,\gamma^\nu\Big]\,\Delta_0(P)\, \G_{\mu\nu}(Q),\qquad P=K-Q\,.
\end{equation}
Evaluating the trace over $\gamma$ matrices one obtains
\begin{equation}\label{eq:sigma_trace}
\bar{\Sigma}(K)
=-4\,g^2C_f\,\sumint_{Q}\;\Big[K^\mu\,P^\nu+P^\mu\,K^\nu-K\cdot P\,g^{\mu\nu}\Big]\,\Delta_0(P)\, \G_{\mu\nu}(Q)\,.
\end{equation}

In order to pursue the calculation, we need to specify a gauge.
We shall use the Coulomb gauge from section \ref{sec:selfenergy}. 
By plugging $\G_{\mu \nu}(Q)$ from Eq.~(\ref{eq:gluon_prop}) into Eq.~(\ref{eq:sigma_trace}), we obtain two components to $\bar{\Sigma}$:
\begin{eqnarray}\label{eq:traces2}
\bar{\Sigma}_{\rm L}(K) & = & -4\,g^2C_f\,\sumint_{Q}\;\Big[\kOEb \,\pOE+\k \cdot \p \Big]\,\Delta_0(P)\, G_{\rm L}(Q)\,,\nonb\\
\bar{\Sigma}_{\rm T}(K) & = & -8\,g^2C_f\,\sumint_{Q}\;\Big[\kOEb\,\pOE-(\hat\q \cdot \k )(\p \cdot \hat\q )\Big]\,\Delta_0(P)\,G_{\rm T}(Q)\,,
\end{eqnarray}
where $\kOE$, $\qOE$ and $\pOE$ are evaluated at the Matsubara frequencies $\kOE = i \omega$, $\qOE = i \omega_n$ and $\pOE = i \omega_m = i\omega - i\omega_n$.

\subsection{Matsubara sums and thermal contributions}\label{sec:fermion_Matsubara}
The contributions to $\Sigma$ that we labelled as ``thermal''
are those which after performing the Matsubara sum in (\ref{eq:traces2}) 
contain a thermal factor $n(|p_0|)$
or $f(|p_0|)$ which vanishes when $T\rightarrow 0$. 
Note that these contributions do not account for the complete temperature dependences
of $\Sigma$, since part of the latter is implicit in the dressed
gluon propagator. The Matsubara sums are performed in Appendix \ref{sec:Matsubara_fermion}.
The result is:
\begin{eqnarray}
\bar{\Sigma}_{\rm L,\,th}(K) & = & -\frac{g^2C_f}{\pi^2}\,\int_0^\infty dp\,p\,f_p\,\left[1+\frac{k^2+p^2}{4kp}\log\left(\frac{k-p}{k+p}\right)^2\right]\nonb\\
& - & 4\,g^2C_f\,\intP \;\Big[\kOEb \,p_0+\k \cdot \p \Big]\,\sigma_0(P)\, G_{\rm L}(K-P)\nonb\\
& + & 4\,g^2C_f\,\intQ \;\Big[\kOEb \,(\kOE -q_0)+\k \cdot \p \Big]\,\Delta_0(K-Q)\, \sigma_{\rm L}(Q)\,,\\
\bar{\Sigma}_{\rm T,\,th}(K) & = & -8\,g^2C_f\,\intP \;\Big[\kOEb \,p_0-(\hat\q \cdot \k )(\p \cdot \hat\q )\Big]\,\sigma_0(P)\, G_{\rm T}(K-P)\nonb\\
& + & 8\,g^2C_f\,\intQ \;\Big[\kOEb \,(\kOE -q_0)-(\hat\q \cdot \k )(\p \cdot \hat\q )\Big]\,\Delta_0(K-Q)\, \sigma_{\rm T}(Q)\,,
\end{eqnarray}
with $\sigma_0(P)=\epsilon(p_0)\,f_{|p_0|}\,\rho_0(P)$ and $\sigma_{\rm L,\,T}(Q)=\epsilon(q_0)\,n_{|q_0|}\,\rho_{\rm L,\, T}(Q)$ and 
$K^\mu = (i \omega_l,\k)$. These integrals are UV finite 
thanks to the presence of the thermal factors.
The continuation to the real axis $k_0 = i\omega_l \rightarrow k_0=\omega+i\epsilon$ and the light-cone limit $\omega \rightarrow k$ pose no problems, since the integrand is always well defined. After performing the angular integrals analytically,
one can perform the remaining $2$-dimensional integrals numerically.

Since integrations in $\bar\Sigma_{\rm th}$ are effectively
cut off by the temperature, the necessity of introducing an
ultraviolet cutoff below the scale of the Landau pole is no
practical problem for temperatures $T\ll \Lambda_L$. The results
will be independent on how such a cutoff is introduced. However,
this will be different for the remaining ``vacuum'' contributions.


\subsection{Vacuum contributions and renormalization}\label{sec:fermion_vac}

The remaining contributions $\Sigma_{\rm vac}$,
which contain temperature dependences (only)
through the spectral data of the gluon propagator,
can be computed by replacing the discrete sum in (\ref{eq:traces2}) by a continuous integral, yielding a $4$-dimensional Euclidean integral:
\begin{eqnarray}\label{SigmaVac4D}
\bar{\Sigma}_{\rm L}^{\rm vac}(K) & = & -4\,g^2C_f\,\intQE \;\Big[\kOEb \,\pOEb +\k \cdot \p \Big]\,\Delta_0(P)\, G_{\rm L}(Q)\,,\nonb\\
\bar{\Sigma}_{\rm T}^{\rm vac}(K) & = & -8\,g^2C_f\,\intQE \;\Big[\kOEb \,\pOEb -(\hat\q \cdot \k )(\p \cdot \hat\q )\Big]\,\Delta_0(P)\,G_{\rm T}(Q)\,,
\end{eqnarray}
where $K=(k_{0},k)=(i\omega_l ,\k)$, $P=(p_{0},\p)=(i\omega_{m},p)$,
and $Q=(q_{0},q)=(i\omega_{n},q)$ have imaginary frequency 
to start with, and $P+Q=K$. 
Since our theory
contains a Landau pole, we have to introduce a cutoff below that scale,
and it is important that this cutoff is implemented in
an Euclidean invariant way even when the result would
be finite in dimensional regularization. This is not only important for quantities affected by the Landau pole but also for quantities free from the Landau singularity, as we illustrate in what follows.

\subsubsection{Cutoff implementation example}\label{sec:cutoffexample}

To give an example 
of difficulties encountered with a non Euclidean-invariant cutoff,
consider the expression $J(K)=I(K)-I(0)$ with
\begin{equation}
I(K)\equiv \intQE \frac{1}{(Q-K)^{2}}\, .
\end{equation}
This quantity appears in the calculation of $\bar{\Sigma}_{\rm vac}[G_0](K)$ and we would like to evaluate it on the light cone $K^2=0$. In dimensional regularization one can shift
the integration momentum, and the result is $[I(K)-I(0)]\rightarrow [I(0)-I(0)]=0$ for arbitrary Euclidean momentum $K$. Hence, the
light-cone value of $J$ is zero. Is it possible to recover this result by using an explicit cutoff?

Let us first introduce a cutoff that violates Euclidean invariance: a cutoff
that only applies to three-dimensional momenta, not to the frequency
(note that we assume imaginary values for $q_{0}$ and $k_{0}$ such
that the integrand does not contain any poles along the integration path):\begin{eqnarray}
I(K) & = & \frac{2\pi}{(2\pi)^{4}}\int_{-i\infty}^{i\infty}\frac{dq_{0}}{i}\int_{0}^{\Lambda}dq\int_{-1}^{1}d(\cos\theta)\frac{q^{2}}{(q_{0}-k_{0})^{2}-(q^{2}+k^{2}-2qk\cos\theta)}\nonumber \\
 & = & -\frac{1}{8\pi^{2}}\Lambda^{2}+\frac{1}{24\pi^{2}}k^{2}.\end{eqnarray}
The integrations over $\theta$, $q_0$, and $q$ are straightforward (in this order), but 
the result we obtain, $I(K)-I(0) = k^2/{24\pi^2}$, is clearly in contradiction
to the result of dimensional regularization. This result turns out to be independent of $k_{0}$ and can thus trivially be continued to Minkowski space leading to a non-zero value on the light cone. It turns out that this is a consequence
of having introduced a cutoff procedure that violates Euclidean rotation invariance for the expression $I(K)$.

Thus, if one has to split frequency
and momentum, one should still do it in a Euclidean invariant
way:\begin{eqnarray}
I(K) & = & \frac{2\pi}{(2\pi)^{4}}\int_{0}^{\Lambda}dq\int_{-i\sqrt{\Lambda^{2}-q^{2}}}^{i\sqrt{\Lambda^{2}-q^{2}}}\frac{dq_{0}}{i}\int_{-1}^{1}dx\frac{q^{2}}{(q_{0}-k_{0})^{2}-(q^{2}+k^{2}-2qkx)}\nonumber \\
 & = & \frac{\pi}{(2\pi)^{4}}\frac{1}{k}\int_{0}^{\Lambda}dq\,q\int_{-i\sqrt{\Lambda^{2}-q^{2}}}^{i\sqrt{\Lambda^{2}-q^{2}}}\frac{dq_{0}}{i}\log\frac{(q_{0}-k_{0})^{2}-(q-k)^{2}}{(q_{0}-k_{0})^{2}-(q+k)^{2}}.\label{IK4D}\end{eqnarray}
The angular integration along $x=\cos\theta$ poses no problem for
imaginary $q_{0}=i\omega_{n}$ and $k_{0}=i\omega_{l}$ as the integrand
is finite and continuous for all $-1\leq x\leq1$. 
The integration
along $q_{0}$ is more problematic as the integrand
\begin{equation}
\log\frac{(q_{0}-k_{0})^{2}-(q-k)^{2}}{(q_{0}-k_{0})^{2}-(q+k)^{2}}
\label{integrand}
\end{equation}
has singularities at $q_{0}=k_{0}\pm|q\pm k|$ 
and branch cuts
which might interfere with the integration path for an external Minkowski $K^\mu=(\omega,k)$.
The particular
form of writing (\ref{integrand}) this way (i.e.~not splitting up
the logarithm $\log(X/Y)$ into $\log X-\log Y$) leads to an analytic
structure in the $q_{0}$ complex plane as shown in Fig.~\ref{cap:1}:
\begin{figure}
\begin{center}\includegraphics[%
  scale=0.8]{
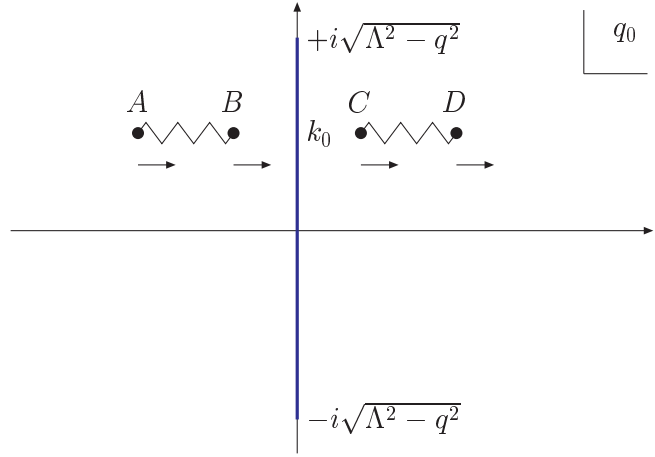}\end{center}

\caption{The branch cut structure for the $q_{0}$ integration in (\ref{IK4D})
with the logarithmic singularities at $A=k_{0}-|q+k|$, $B=k_{0}-|q-k|$,
$C=k_{0}+|q-k|$, $D=k_{0}+|q+k|$. For purely imaginary $k_{0}$
(in the plot just between $B$ and $C$), the integration path runs
from $-i\sqrt{\Lambda^{2}-q^{2}}$ to $i\sqrt{\Lambda^{2}-q^{2}}$
and does not cross any logarithmic branch cut. \label{cap:1}}
\end{figure}
 a branch cut connecting the left two singularities and a branch cut
connecting the rightmost two singularities. Note that if we start with
purely imaginary $k_{0}$, the $q_{0}$ integration poses no problem
since all singularities and branch cuts are away from the imaginary
integration axis (except for $q=k$).

\begin{figure}
\begin{center}\includegraphics[%
  scale=0.8]{
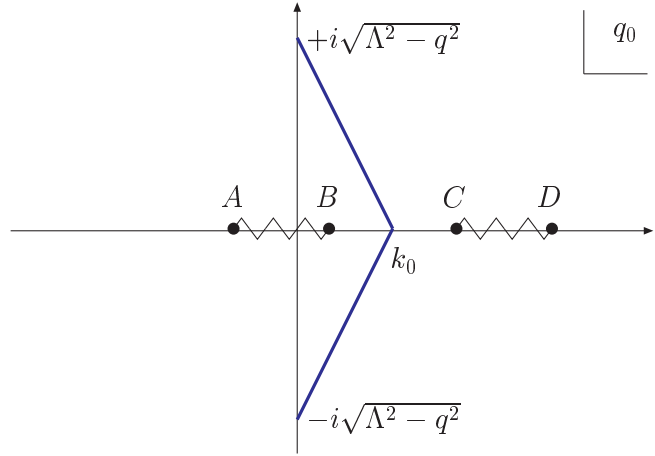}\end{center}

\caption{As the rotation $k_{0}=i\omega\rightarrow\omega$ moves all the singularities
at the same time, we should also deform the integration path in order
to avoid crossing a singularity or branch cut. \label{cap:2}}
\end{figure}
In order to evaluate (\ref{IK4D}) on the light cone, we have to rotate
$k_{0}$ from the imaginary to the real axis $k_{0}=i\omega_l\rightarrow\omega+i\epsilon$.
By doing so, all four singularities will move to the right, and one
of them will eventually cross the integration path when $\Re k_{0}>|q-k|$.
We want $I(K)$ to stay an analytic function. This is only possible
if we avoid singularities and branch cuts by deforming the (numerical) integration
path as shown in Fig.~\ref{cap:2}. A deformation of a complex path
that does not cross poles, singularities, or branch cuts, will not
change the result of a complex integration.

Note that $q=k$ gives a {}``pinch singularity'', that is a point
where in Fig \ref{cap:1} two singularities pinch the integration
path. It turns out that in our case this singular point 
corresponds to a discontinuity of the first derivative of the $q$ integrand, 
but is otherwise harmless for the $q$ integration. Also note that when $q>2k=2k_0$, we do not have
to deform the integration path as the point $B=k_0-|q-k|<0$ in Fig.~\ref{cap:2} stays on the left side of the integration path even on the light cone. We can choose the integration path such that all path deformation is performed for $Q_E=\sqrt{-q_0^2+q^2}<\Lambda_1$ and no path deformation needed for $Q_E>\Lambda_1$ as long as the intermediate cutoff $\Lambda_1>2k$.

Performing the numerical integration with the modified path indeed gives the desired result $I(K) - I(0) = 0$ 
also on the light cone,
in accordance with dimensional regularization. 
Blindly integrating across the branch cut without deforming the integration path would give a wrong, non-vanishing result on the light cone.

\subsubsection{Full cutoff implementation}

\begin{figure}
\begin{center}\includegraphics[%
  scale=0.8]{
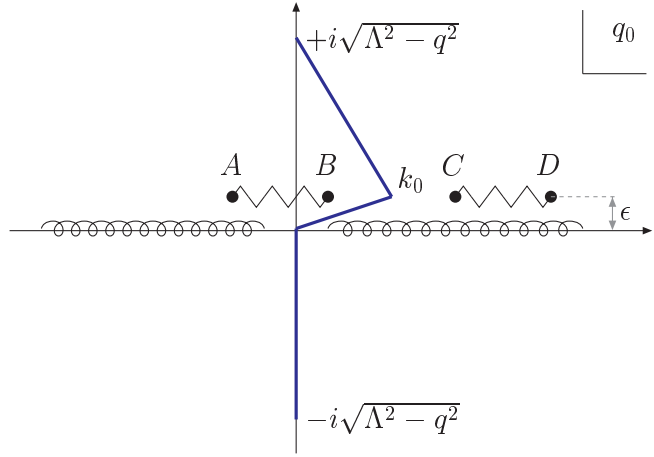}\end{center}

\caption{A general propagator with self-energy $\Pi$ will introduce additional
cuts and poles along the real $q_{0}$ axis. In this case, the integration
has to go through the origin and $k_{0}$ has to be shifted slightly
off the real axis to $k_{0}=\omega+i\epsilon$. \label{cap:3}}
\end{figure}
We can now perform the angular integration in the vacuum contributions (\ref{SigmaVac4D}) and obtain

\begin{eqnarray}\label{SigmaVac4Dangular}
\bar{\Sigma}_{\rm L}^{\rm vac}(K) & = & -4\,g^2C_f\,
\frac{2\pi}{(2\pi)^{4}}
\int_{0}^{\Lambda}dq\;q^2
\int_{-i\sqrt{\Lambda^{2}-q^{2}}}^{i\sqrt{\Lambda^{2}-q^{2}}}\frac{dq_{0}}{i} \; 
G_{\rm L}(Q)\nonb\\
 & \times & \left[1-\frac{(q_0-k_0)(q_0-3k_0)-q^2+k^2}{4k q}
\log\frac{(q_{0}-k_{0})^{2}-(q-k)^{2}}{(q_{0}-k_{0})^{2}-(q+k)^{2}}
\right] \,,\nonb\\
\bar{\Sigma}_{\rm T}^{\rm vac}(K) & = & -8\,g^2C_f\,
\frac{2\pi}{(2\pi)^{4}}
\int_{0}^{\Lambda}dq\;q^2
\int_{-i\sqrt{\Lambda^{2}-q^{2}}}^{i\sqrt{\Lambda^{2}-q^{2}}}\frac{dq_{0}}{i} \; 
G_{\rm T}(Q) \;
\left[-\frac{(q_0-k_0)^2+q^2-k^2}{2q^2}\right.\nonb\\
 & + & \left.\frac{\left((q_0-k_0)^2-k^2\right)^2-q^2\left(4k_0(k_0-q_0)+q^2\right)}{8k q^3}
\log\frac{(q_{0}-k_{0})^{2}-(q-k)^{2}}{(q_{0}-k_{0})^{2}-(q+k)^{2}}
\right]\,.
\end{eqnarray}
The bare propagator
$\Delta_{0}(P)=\Delta_{0}(K-Q)$ gives the same logarithmic expression of the
form (\ref{integrand}) as in our example. But $G_{{\rm L}}(Q)$ and $G_{{\rm T}}(Q)$,
which contain the gluon self-energies $\Pi_{{\rm L}}$ and $\Pi_{{\rm T}}$,
provide additional branch cut structures along the real $q_{0}$ axis
such that the path deformation is further restricted as shown in 
Fig.~\ref{cap:3}. The integration path goes from $q_{0}:\,-i\sqrt{\Lambda^{2}-q^{2}}\rightarrow0\rightarrow k_{0}(=\omega+i\epsilon)\rightarrow i\sqrt{\Lambda^{2}-q^{2}}$. 

Another numerical subtlety is involved: Keeping $\epsilon$ small
but fixed will limit the upper integration bound of the $q_{0}$ integration
to $i\sqrt{\Lambda^{2}-q^{2}}>i\epsilon$ (see Fig.~\ref{cap:3})
which means that the $q$-integration is limited to  $0\leq q<\sqrt{\Lambda^{2}-\epsilon^{2}}$
. The error introduced by this turns out to be at least of the order
$O(\epsilon^2)$.

\subsection{Renormalization}\label{sec:ren}

\subsubsection{Large-$N_f$ limit} 
As already mentioned, the disappearance of the counterterms $\delta Z_\psi$ and $\delta M$ from the entropy formula indicates that $\Re \bar{\Sigma}_{\rm vac}$ has to be finite on the light cone. Since the propagator $G$ used to compute this quantity contains a Landau pole, we have to be more specific about what we call ``finite''. $\Re \bar{\Sigma}_{\rm vac}$ is computed with an explicit Euclidean invariant cutoff $\Lambda\ll \Lambda_{\rm L}$. The finiteness of $\Re \bar{\Sigma}_{\rm vac}$ means that it is insensitive to $\Lambda$ in a broad region of momenta $T\ll \Lambda\ll \Lambda_{\rm L}$. In order to understand this more explicitly, we can use the linearity of $\Re \bar{\Sigma}_{\rm vac}$ with respect to $G$ in order to write $\Re \bar{\Sigma}_{\rm vac}=\Re \bar{\Sigma}_{\rm vac}[G_{\rm vac}]+\Re \bar{\Sigma}_{\rm vac}[G-G_{\rm vac}]$. Using the asymptotic behaviour of $\Pi_{\rm th}$, one can check that the second term of this expression is insensitive to $\Lambda$. Thus we only need to check the finiteness of $\Re \bar{\Sigma}_{\rm vac}[G_{\rm vac}]$ on the light cone.

As explained in the previous section, the use of an Euclidean invariant cutoff allows one to obtain the same results that one would obtain in dimensional regularization if there were no Landau pole. If we suppose that there is no Landau pole, then it is possible to show that $\Re \bar{\Sigma}_{\rm vac}[G_{\rm vac}]=0$ on the light cone. Indeed this is a gauge invariant quantity and one can compute it in any gauge. Choosing a covariant gauge, the cancellation is straightforward according to covariance and dimensional regularization. Of course we cannot rigorously use dimensional regularization because of the Landau pole. However using an explicit Euclidean invariant cutoff, it is possible to check that $\Re \bar{\Sigma}_{\rm vac}[G_{\rm vac}]=0$ on the light cone. 

To do so, we start from equations (\ref{SigmaVac4Dangular}) 
and transform the integration variables to 4-dimensional variables according to
\begin{equation}
\int_{0}^{\Lambda}dq\;q^2
\int_{-i\sqrt{\Lambda^{2}-q^{2}}}^{i\sqrt{\Lambda^{2}-q^{2}}}\frac{dq_{0}}{i} \; 
f(q_0,q)
=
\int_{0}^{\Lambda}dQ_E\;Q_E^3
\int_{0}^{\pi}d\xi\;\sin^2 \xi \; 
f(i Q_E \cos \xi, Q_E \sin \xi)\;.
\end{equation}
The integrand can then be expanded for large $Q_E$ 
and, since in the free case $G_{\rm T}(Q_E) = G_{\ell}(Q_E) = -G_{\rm L}(Q_E)/\sin^2 \xi = (Q_E^2 + \Pi_{\rm vac}(Q_E^2))^{-1}$ is independent of $\xi$, we can perform the $\xi$ integration and obtain
\begin{eqnarray}\label{SigmaVac4Dexpanded}
\Re \bar{\Sigma}_{\rm L}^{\rm vac}(K) & = & -4\,g^2C_f\,
\frac{2\pi}{(2\pi)^{4}}
\int_{0}^{\Lambda}dQ_E\;Q_E^3 
G_{\ell}(Q_E)\nonb\\
 & \times & \frac{\pi}{3} \left[
-\frac{4k^2}{Q_E^2}
+\frac{8k^4}{5Q_E^4}
-\frac{4k^2(9k^4+42k^2k_0^2-35k_0^4)}{35Q_E^6}+O\left(\frac{1}{Q_E^8}\right)
\right] \,,\nonb\\
\Re \bar{\Sigma}_{\rm T}^{\rm vac}(K) & = & -8\,g^2C_f\,
\frac{2\pi}{(2\pi)^{4}}
\int_{0}^{\Lambda}dQ_E\;Q_E^3
G_{\rm T}(Q_E)\nonb\\
 & \times & \frac{\pi}{6} \left[
\frac{k^2+3k_0^2\pi}{Q_E^2}
-\frac{8k^4\pi}{5Q_E^4}
+\frac{8(k^6+7k^4k_0^2)\pi}{35Q_E^6}+O\left(\frac{1}{Q_E^8}\right)
\right]\,.
\end{eqnarray}
The first term of each expansion is logarithmically divergent if transverse and longitudinal parts are taken separately. Only when adding the components together to form $\bar{\Sigma}^{\rm vac}(K) = \bar{\Sigma}_{\rm L}^{\rm vac}(K) + \bar{\Sigma}_{\rm T}^{\rm vac}(K)$, the logarithmic divergence vanishes on the light cone $k_0^2=k^2$:
\begin{eqnarray}\label{SigmaVac4Dadded}
\Re \bar{\Sigma}^{\rm vac}(K) & = & -4\,g^2C_f\,
\frac{2\pi^2}{(2\pi)^{4}}
\int_{0}^{\Lambda}dQ_E\;Q_E^3 
G_{\rm vac}(Q_E)\nonb\\
 & \times & \left[
\frac{k_0^2-k^2}{Q_E^2}
+\frac{4k^2(k_0^2-k^2)(k^2+5k_0^2)}{15Q_E^6}+O\left(\frac{1}{Q_E^8}\right)
\right] \,.
\end{eqnarray}

\subsubsection{HTL approximation}
Unlike the case of large $N_f$, the disappearance of the counterterms $\delta Z_\psi$ and $\delta M$ from the entropy formula cannot be taken as an indication of the finiteness of $\Re \bar{\Sigma}_{\rm vac}[\hat{G}]$. This is because $\hat{G}$ is built out of approximations of Feynman diagrams whose asymptotic behaviour is changed with respect to the large-$N_f$ case. If one writes $\Re \bar{\Sigma}_{\rm vac}[\hat{G}]=\Re \bar{\Sigma}_{\rm vac}[G_0]+\Re \bar{\Sigma}_{\rm vac}[\hat{G}-G_0]$ one is able to check that the first term of this expression is zero on the light cone (using an Euclidean invariant cutoff). 
For the second term, since the asymptotic behaviour of $\hat{\Pi}$ is different from that of the large-$N_f$ result $\Pi_{\rm th}$, we need a direct check.
In principle one can perform this check as in the previous section, just Eq.~(\ref{SigmaVac4Dexpanded}) would have to be changed when using HTL propagators, since $\hat G_{\rm T}(Q_E,\xi)$ and $\hat G_{\ell}(Q_E,\xi)$ now also depend on the angle $\xi$. One can easily convince oneself that the angular dependence does not affect the leading logarithmic divergence though, as both transverse and longitudinal HTL propagators can be expanded in the form
\begin{equation}
\hat G(Q_E,\xi)=\frac{1}{Q_E^2} 
\left\{
1 + f(\xi)  \frac{m_D^2}{Q_E^2} 
+O\left(\frac{m_D^4}{Q_E^4}\right)
\right\}\;.
\end{equation}
The additional angular dependence $f(\xi)$ is suppressed by a factor $m_D^2/Q_E^2$ at large $Q_E$ and therefore does not influence the leading logarithmic divergent behaviour of Eqs.~(\ref{SigmaVac4Dexpanded}) and (\ref{SigmaVac4Dadded}): the leading divergence still vanishes on the light cone.

\section{Numerical results}\label{sec:results}

\subsection{Numerical implementations}

The evaluation of equation (\ref{SigmaVac4D}) requires three consecutive
numerical integrations (of which one is given by the complex path
described earlier) in order to obtain one value of $\Sigma(\omega=k;\geff^{2})$.
With our implementation this takes of the order of 10 hours on a current
PC (3GHz). This has to be integrated over to obtain the fermionic
contribution to the entropy $S_{{\rm fermion}}(\geff^{2})$ for one
point of $\geff^{2}$. 
Part of the code was therefore ported to run in parallel and final results could be obtained on the ECT{*} Teraflop
Cluster within a couple of weeks that would have taken years on a standard PC.

It turns out that the numeric cancellation in the integrand (\ref{SigmaVac4D})
for large $q$ and (imaginary) $q_{0}$ works best if one integrates
over 4-dimensional Euclidean spheres at fixed $Q_E=\sqrt{-q_{0}^{2}+q^{2}}$
first. No path deformation is needed as long as $Q_E>2k$ as was discussed at the end of section \ref{sec:cutoffexample}. Unfortunately,
the path deformation for 4-dimensional angular variables turns out
to be much more involved (the corresponding deformed path could become
infinitely long due to inverse trigonometric functions), such that
a hybrid approach seems a good compromise between simple path-deformation
for small $Q_E$ and good convergence properties for larger $Q_E$:\begin{eqnarray}
\int_{0}^{\Lambda}d^{4}q & = & \int_{0}^{\Lambda_{1}}d^{4}q+\int_{\Lambda_{1}}^{\Lambda}d^{4}q\nonumber \\
 & = & \int_{0}^{\Lambda_{1}}dq\, q^{2}\int_{-i\sqrt{\Lambda_{1}^{2}-q^{2}}}^{i\sqrt{\Lambda_{1}^{2}-q^{2}}}\frac{dq_{0}}{i}\int d\Omega_{2}+\int_{\Lambda_{1}}^{\Lambda}dQ_E\,Q_E^{3}\int d\Omega_{3}\end{eqnarray}
where $\Lambda_{1}$ is an intermediate cutoff to be chosen such that $2k<\Lambda_{1}\ll\Lambda$
(in our implementation we chose $\Lambda_{1}=2.1k$), and $\int d\Omega_{2}$
and $\int d\Omega_{3}$ denote the 3- and 4-dimensional angular
integrations. Practically, yet another cutoff $\Lambda_{1}<\Lambda_{2}<\Lambda$
is introduced, above which the angular integration $\int d\Omega_{3}$
is performed analytically on the high-temperature series expansion.
This series expansion is also used to show that neither large-$N_{f}$
nor HTL gluon self-energies contain logarithmically divergent
pieces for high momentum shells.

In order to correctly integrate all peaks along the complex path,
the routine for complex path integration is enhanced by providing
information about analytically or numerically known positions in the
complex plane of poles of the propagator, $\omega_{{\rm L}}(q)$ and
$\omega_{{\rm T}}(q)$ \cite{Blaizot:2005fd},
singularities of the logarithms and other
singular points. The integration path is divided into smaller segments
in the vicinity of such points where the integrand can rapidly change
its value by orders of magnitude.

\subsection{Asymptotic thermal quark masses}
\label{sect:resmas}

\begin{figure}
\begin{center}\includegraphics{
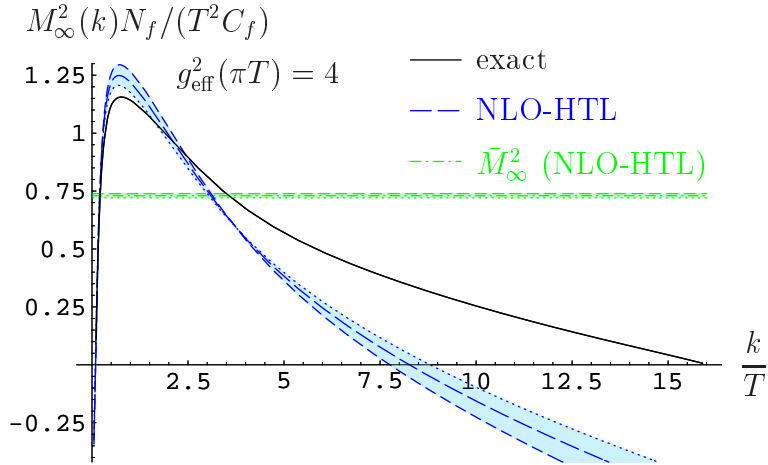}\end{center}
\caption{Asymptotic thermal quark mass squared
(real part of the fermionic self-energy on the light-cone) 
as a function of $k/T$ for $\geff^{2}(\pi T)=4$.
The exact large-$N_f$ 
result is compared to the NLO-HTL calculation and its relevant
average value. The renormalization scale $\muMS$ is varied around
the FAC-m scale by factors of 2 (see text). \label{fig:sigma4}}
\end{figure}

\begin{figure}
\begin{center}\includegraphics{
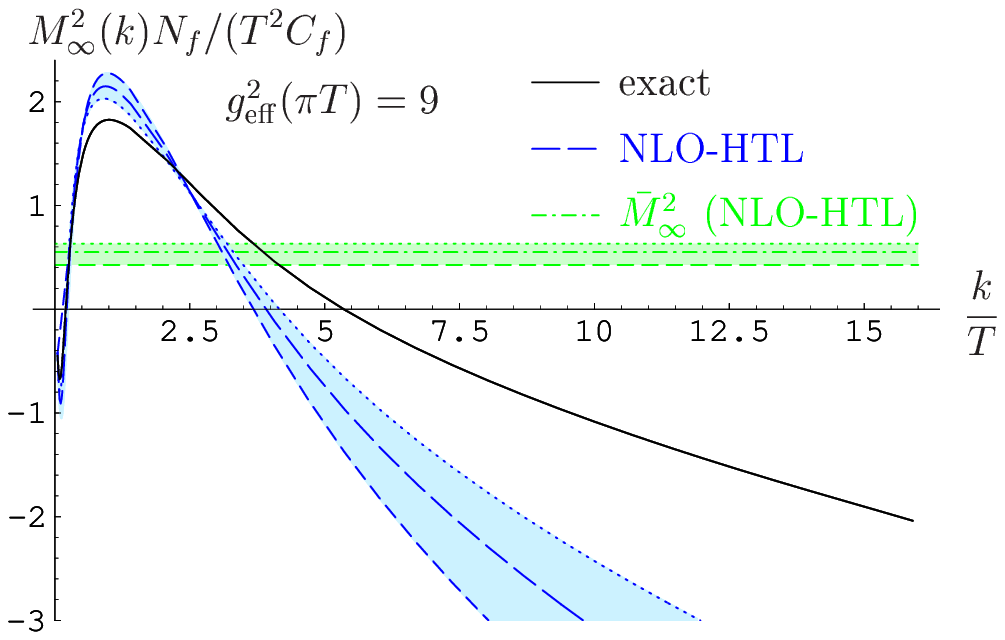}\end{center}
\caption{Same as Fig.~\ref{fig:sigma4} for $\geff^{2}(\pi T)=9$. \label{fig:sigma9}}
\end{figure}

\begin{figure}
\begin{center}\includegraphics{
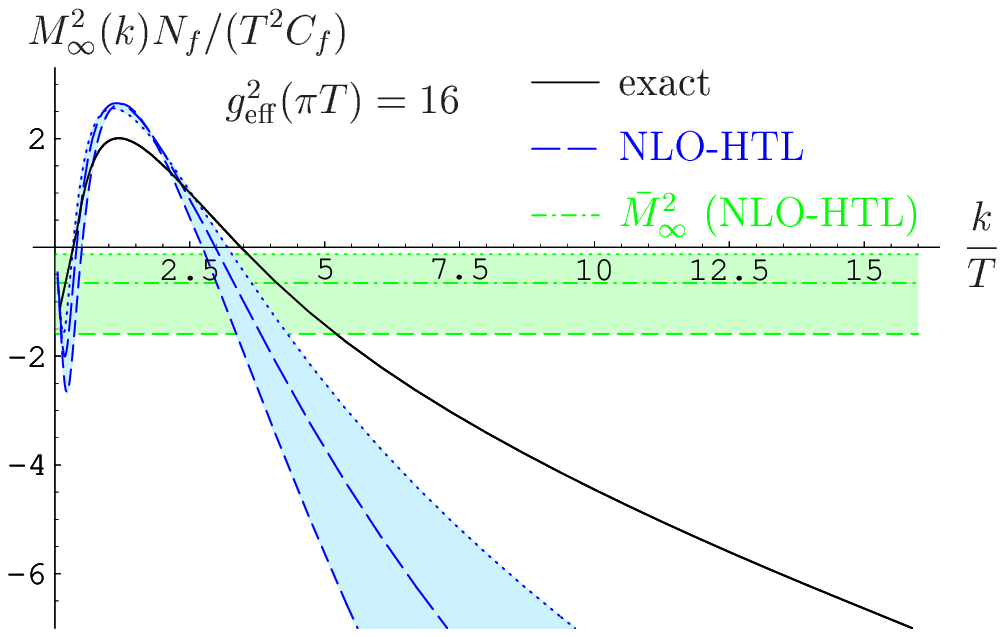}\end{center}
\caption{Same as Fig.~\ref{fig:sigma4} for $\geff^{2}(\pi T)=16$ . \label{fig:sigma16}}
\end{figure}

Figures \ref{fig:sigma4}, \ref{fig:sigma9}, and \ref{fig:sigma16} show 
the results of a numerical calculation of the asymptotic thermal quark
mass squared $M_\infty^2(k)\equiv 2k\Re\Sigma(k_0=k)$ 
for three different values of the coupling $\geff^2(\muMS\!=\!\pi T) = $ 4, 9, and 16,
and normalized by $T^2C_f/N_f$. (Recall that
this quantity is of order $N_f^{-1}$, and it contributes to the
entropy only because there are $N_f$ fermions.)
The exact (nonperturbative) result obtained in the large-$N_f$ limit
is given by the full lines.

As anticipated, the asymptotic thermal quark mass squared is a nontrivial
function of momentum. It is maximal for $k\sim T$ and decays for
large $k\gg T$. The fact that it vanishes and even becomes negative
at very small momentum is not to be taken seriously. 
$M_\infty(k)$ is referred to as asymptotic thermal quark mass, because
only at hard $k\gtrsim T$ does this have the interpretation of
a quasiparticle mass; for smaller $k$ one would have to search
for a self-consistently determined pole of the dressed quark propagator
rather than evaluate the quark self-energy on the tree-level mass
shell, i.e.\ the light-cone. $M^2_\infty(k)$ however also becomes
negative at very large momenta, $k\gg T$. This means that
the dispersion law for hard fermionic modes eventually turns
space-like, which does not necessarily
signal an inconsistency of the theory.
In fact, the group velocity
remains smaller than the speed of light for all $k\gtrsim T$.
Negative $M^2_\infty(k)$ however opens up the possibility
for Cerenkov radiation and thus means that at the corresponding
values of $k$ there is
a qualitative difference to the leading-order
result $\hat M_\infty^2={1\02}\g^2 C_f T^2/N_f$.

The next-to-leading order result as obtained in HTL-resummed
perturbation theory (i.e., replacing the full gauge boson propagator
in the quark self-energy by its HTL approximation)
is given by the dashed lines in
Figs.\ \ref{fig:sigma4}--\ref{fig:sigma16}. 
Qualitatively, the HTL result is similar to the exact large-$N_f$ result.
Quantitatively,
like every perturbative
result, it has a dependence on the renormalization scale.
The HTL result is displayed for a range of renormalization
scales $\muMS$ centered at 
$\bar\mu_{\rm FAC-m}= \exp(\2-\gamma_{\rm E})\pi T \approx 0.926 \pi T$
which is the scale of fastest apparent convergence with respect to the effective mass parameter $m_{\rm E}$ in dimensional reduction 
\cite{Braaten:1996jr}.
The upper (dotted) limiting lines of the shaded bands
correspond to $\2 \times \bar\mu_{\rm FAC-m}$, while the lower (dashed) limiting lines correspond to $2 \times \bar\mu_{\rm FAC-m}$.
The renormalization scale dependence is seen to increase with
the coupling $\g^2$, and also in the domain where $k\gg T$.
For large values of $k$ the HTL result
for the asymptotic mass squared turns more quickly negative than the
full result, but in fact at large values of $k$ the modes
are severely suppressed by the Bose-Einstein distribution factor.
The range $k\sim T$ is more relevant physically (in particular
for the calculation of the entropy to be discussed presently), and
here the agreement with the exact large-$N_f$ result is much better.

\begin{figure}
\begin{center}\includegraphics{
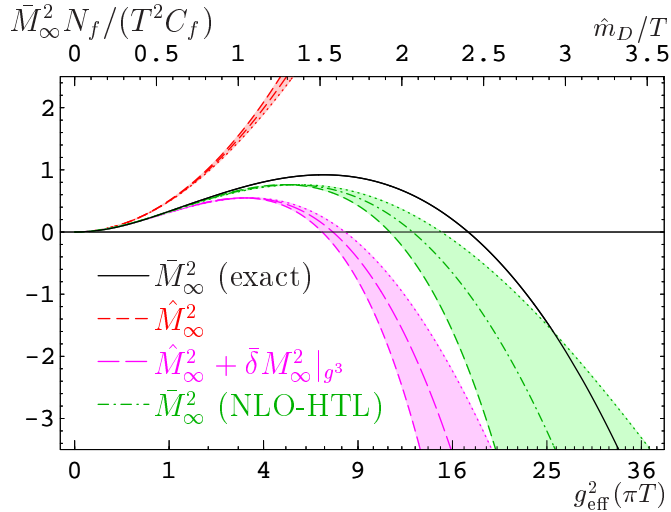}\end{center}
\caption{Comparison of the averaged asymptotic thermal quark mass squared,
$\bar M_{\infty}^{2}$, in various approximations as explained in the text. The renormalization
scale $\muMS$ is varied around the FAC-m scale by factors of 2. 
\label{fig:minfinity}}
\end{figure}

For the computation of the fermionic contribution to the entropy,
the asymptotic thermal mass
appears in form of the weighted average given in Eq.~(\ref{Minftybarfull}).
This averaged asymptotic thermal mass is given for the
HTL approximation by the flat horizontal bands labelled
$\bar M_\infty^2$ (NLO-HTL). 
For the entire range of couplings that we can consider without
becoming sensitive to the scale of the Landau pole, the
averaged asymptotic quark masses are compared in
Fig.\ \ref{fig:minfinity}. The full line is again
the exact large-$N_f$ result, given as a function
of $\g^2(\pi T)$. The short-dashed lines correspond to
the leading-order HTL value $\hat M_\infty^2$ with
renormalization scale varied as before.
The perturbative next-to-leading order result
(\ref{Minftypth}) is given by the longer-dashed lines.
Finally the full NLO-HTL-resummed result, not truncated at order $\g^3$,
is given by the dash-dotted lines.

As one can see in Fig.\ \ref{fig:minfinity}, the 
NLO-HTL-resummed result represents a considerable improvement
over the HTL-resummed result truncated at order $\g^3$
for $\g^2\gtrsim 4$, which corresponds to $\hat m_D/T \gtrsim 1$.

\subsection{Numerical results for the entropy
and discussion}

\begin{figure}
\begin{center}\includegraphics{
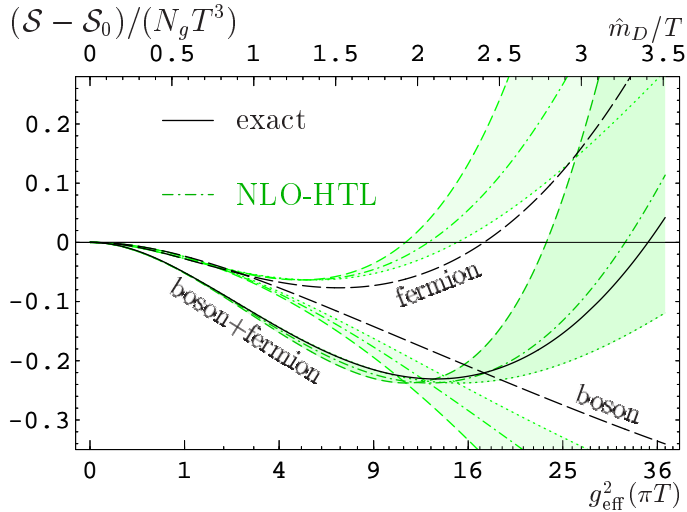}\end{center}
\caption{Entropy in the large-$N_{f}$ limit
separated into a bosonic and a fermionic part, corresponding respectively to 
the first and second line of Eq.~(\ref{Slnf}). 
The NLO-HTL approximation
to the two parts of the
entropy is depicted in dashed lines, where the renormalization
scale $\muMS$ is varied around the FAC-m scale by factors of 2. 
The combined
NLO-HTL result shows remarkable agreement with the exact large-$N_{f}$
result for all couplings with HTL Debye mass 
$\hat m_D \lesssim 2.5 T$.
\label{fig:entropycontributions}}
\end{figure}

\begin{figure}
\begin{center}\includegraphics{
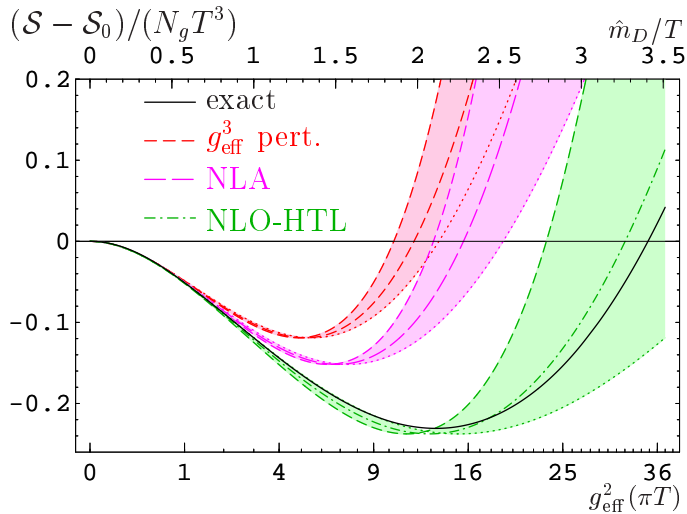}\end{center}
\caption{Entropy in the large-$N_{f}$ limit, comparing the exact large-$N_{f}$
result to the strictly perturbative expansion through order $\geff^{3}$,
the NLA result, and the NLO-HTL result. The renormalization scale
$\muMS$ is varied around the FAC-m scale by factors of 2. \label{fig:entropy}}
\end{figure}

Figure \ref{fig:entropycontributions} 
shows the numerical results of the entropy
calculation. The full line is the entropy density ${\cal S} =
\left(\partial{P}/\partial T \right)_\mu$ as it has been obtained
earlier \cite{Moore:2002md,Ipp:2003zr,Ipp:2003jy} by a numerical
derivative of the pressure $P$ from Eq.~(\ref{eq:PNLO}) for massless
fermions $m=0$. It is a very non-trivial numerical test that the
result for the entropy as obtained from the $\Phi$-derivable two-loop
approximation, Eq.~(\ref{S2loop}), reproduces exactly the same result
in the large-$N_f$ limit, as has been discussed in section
\ref{sec:largeNf2PI}. Indeed, only after implementing the correct path
deformation for the calculation of $\Sigma$, we have been able to
reproduce the entropy to three or four digits, where the accuracy was
limited only by the high calculation cost for the multi-dimensional
integrals. In Figure~\ref{fig:entropycontributions} 
the two curves lie perfectly on
top of each other for the whole range of couplings displayed.

We show the results in a range of couplings $\geff^2 (\pi T) \lesssim 36$ where the influence of the Landau pole can be neglected. As in \cite{Moore:2002md,Ipp:2003zr,Ipp:2003jy} we obtain differences in the entropy on the percent level for the largest couplings $\geff^2 (\muMS = \pi T)\approx 36$ by varying the numerical cutoff $\Lambda^2 = a \Lambda_{\rm L}^2$ in the range $a = 1/4\, ..\, 1/2$.

According to Eq.~(\ref{Slnf}), the large-$N_f$ entropy as well
as the HTL approximation thereof is composed of a bosonic and a fermionic
contribution. The fermionic one is entirely given by the
weighted average (\ref{Minftybarfull}) of the asymptotic thermal
quark mass that we have discussed above and it is reproduced
in Fig.~\ref{fig:entropycontributions} by the line marked ``fermion'',
together with its NLO-HTL approximation.
The calculation of the
bosonic contribution is less demanding computationally.
The result is shown by the line marked ``boson'' in 
Fig.~\ref{fig:entropycontributions}.
The HTL approximation to the bosonic contribution 
does not require NLO corrections to the asymptotic bosonic masses
in the large-$N_f$ limit, and it has been
calculated completely already in Refs.\ \cite{Blaizot:1999ip,Blaizot:2000fc}.
Evaluated for the range of renormalization scales considered here
it gives the band below the line marked ``boson''.
Perhaps fortuitously, the errors of the bosonic contribution
to the HTL approximation of the entropy are opposite in sign
from those of the fermionic contribution. The sum total turns out
to reproduce the exact result with astounding accuracy up to $\g^2\sim 16$.
For larger coupling the renormalization scale dependence quickly
becomes enormous, however the central value determined by
the optimized scale $\muMS=\bar\mu_{\rm FAC-m}$ remains
amazingly close to the exact result for all values of $\g$.
In Ref.~\cite{Ipp:2003jy} a similarly successful approximation
was constructed by using optimized renormalization scales
on a perturbative result, which however required to include 
all contributions up to and
including order $\g^6$.

In Fig.~\ref{fig:entropy} the exact and the NLO-HTL result is
also compared to simpler approximations. The short-dashed lines
give the strictly perturbative result up to and including order
$\g^3$, the longer-dashed lines labelled ``NLA'' (next-to-leading approximation) show
the result of comparing
the bosonic HTL entropy with a perturbative
approximation to the averaged asymptotic thermal quark mass
presented before in Ref.~\cite{Rebhan:2003fj}. The latter corresponds
to the large-$N_f$ limit of the HTL-resummed entropies obtained
in Refs.\ \cite{Blaizot:1999ip,Blaizot:1999ap,Blaizot:2000fc}.
While this does lead to an important improvement for couplings
such that $\hat m_D/T \lesssim 1.5$, it is remarkable that
a complete HTL resummation in the asymptotic thermal mass
can push the range where HTL resummation works well up to
$\hat m_D/T \sim 2.5$.

\section{Conclusions and outlook}

We have calculated the large-$N_f$ limit of the entropy
of ultrarelativistic gauge theories by evaluating separately
the contributions from bosonic and fermionic quasiparticles.
Since in the large-$N_f$ limit, the 2-loop $\Phi$-derivable
approximation becomes exact, this allowed us to assess
the error that is being made when fully dressed propagators and
self-energies are replaced by their HTL approximations.

In the case of the bosonic contributions, these approximations
correspond to replacing a full fermion loop in the
dressed gluon propagator by its HTL approximation.
The fermionic contributions to the entropy, on the other hand,
involve the momentum-dependent asymptotic thermal quark mass
where the HTL approximation has to be carried to next-to-leading
order. In both cases the HTL approximation turned out to give
remarkably good results when compared
with the exact large-$N_f$ results, even for fairly large coupling. 
Combining these contributions,
the final result for the entropy in the NLO-HTL approximation
turned out to be amazingly accurate 
up to $\hat m_D/T \sim 2.5$, and even beyond when the 
renormalization scale is fixed by the requirement of
fastest apparent convergence of the electric mass in dimensional
reduction. Its quality is then comparable to optimized 
perturbative results including terms through order $\g^6$.
This seems to support the conclusions of Ref.~\cite{Blaizot:2003iq,Ipp:2003yz}
where optimised dimensional-reduction results for QCD to
order $g^6\log(g)$ were found to agree well with available
lattice data and in turn with the estimates from
the HTL entropy.

The improvement achieved by including the full momentum dependence
of asymptotic thermal masses and keeping all effects of the
resummation of hard thermal loops is encouraging for further
developments of this approach. A straightforward extension of
the present calculations would be the inclusion of finite quark chemical
potential. The calculation of quark densities and quark number susceptibilities
\cite{Blaizot:2001vr,Blaizot:2002xz,Rebhan:2003wn} 
for small-$N_f$ QCD within NLO-HTL resummation
in fact requires exclusively the asymptotic quark thermal masses
that were calculated in the present work for zero chemical potential.
By also calculating
the full momentum dependent asymptotic gluon masses,
one could finally
complete the existing HTL results 
\cite{Blaizot:1999ip,Blaizot:1999ap,Blaizot:2000fc}
for the entropy of full QCD
in the 2-loop $\Phi$-derivable approximation.

\acknowledgments

We would like to thank Edmond Iancu for many fruitful discussions
and collaboration in the earlier stages of the present work.
Urko Reinosa has been supported by the Austrian Science Fund FWF,
project no.\ P16387-N08.


\appendix

\section{Matsubara sums}

\subsection{Gluon self-energy}\label{sec:Matsubara_gluon}
\noindent{The gluon self-energy to leading order in $\Nf$ is given by:}
\begin{equation}
\Pi_{\rm b}^{\mu\nu}(Q)=\g^2 \, \tr \sumintp_{K} \; \, \gamma^\mu\,S_0(K)\,\gamma^\nu\,S_0(P)\,,
\end{equation}
with $P=K-Q$. It is convenient to split this self-energy into a vacuum and a thermal piece $\Pi_{\rm b}=\Pi_{\rm b, vac}+\Pith$ (the label ``b'' stands for ``bare'' and is here to remind that the vacuum piece contains UV divergences). The vacuum piece is simply obtained by replacing the discrete Matsubara sum by a continuous integral, thus leading to a $4$-dimensional Euclidean integral:
\begin{equation}
\Pi_{\rm b,\,vac}^{\mu\nu}(Q)=\g^2 \, \tr\, \intKE  \; \, \gamma^\mu\,S_0(K)\,\gamma^\nu\,S_0(P)\,.
\end{equation}
The thermal piece is obtained after performing the Matsubara sums and separating the thermal dependent pieces. A convenient way to proceed is to introduce the spectral representation (\ref{eq:spectral_fermion}) for each of the fermionic propagators:
\begin{equation}
\Pi_{\rm b}^{\mu\nu}(Q) = \g^2 \, \tr\, \intpO \intkO \intbfk \; \, \gamma^\mu\,(\slashchar{K}-M)\,\rho_0(K)\,\gamma^\nu\,(\slashchar{P}-M)\,\rho_0(P) \frac{1}{\beta}\sum_l\frac{1}{(k_0-i\omega_l)(p_0-i\omega_m)}\,,
\end{equation}
where $\omega_m=\omega_l-\omega_n$ and $q_0 = i\omega_n$. One can now easily perform the Matsubara sum and extract the thermal dependent part:
\begin{equation}
\frac{1}{\beta}\sum_l\frac{1}{(k_0-i\omega_l)(p_0-i\omega_l+i\omega_n)}=\frac{-f(k_0)+f(p_0)}{p_0-k_0+i\omega_n}=\frac{\epsilon(k_0)f(|k_0|)-\epsilon(p_0)f(|p_0|)}{p_0-k_0+i\omega_n}+\mbox{vac}\,,
\end{equation}
where we have used $f(k_0)=\theta(-k_0)-\epsilon(k_0)f(|k_0|)$. The thermal part of the gluon self-energy now reads (we do not use the label ``b'' since this contribution is UV finite):
\begin{equation}
\Pi_{\rm th}^{\mu\nu}(Q) = \g^2 \, \tr\, \intpO \intkO \intbfk \; \, \gamma^\mu\,(\slashchar{K}-M)\,\rho_0(K)\,\gamma^\nu\,(\slashchar{P}-M)\,\rho_0(P)
\frac{\epsilon(k_0)f(|k_0|)-\epsilon(p_0)f(|p_0|)}{p_0-k_0+\qOE }\,.
\end{equation}
Using the spectral representation for the fermion propagator, the parity properties of $\Pi^{\mu\nu}$ together with a straightforward change of variables (which is justified since the cut-off can be sent to infinity in the thermal contribution), one obtains:
\begin{eqnarray}
\Pi_{\rm th}^{\mu\nu}(Q) & = & 2\,\g^2 \, \tr\, \intK  \; \, \gamma^\mu\,(\slashchar{K}-M)\,\sigma_0(K)\,\gamma^\nu\,S_0(K-Q)\,,
\end{eqnarray}
with $\sigma_0(K)=\epsilon(k_0)f(|k_0|)\rho_0(K)$.

\subsection{Fermion self-energy}\label{sec:Matsubara_fermion}
\noindent{In this section, we compute the fermionic Matsubara sums:}
\begin{eqnarray}
\bar{\Sigma}_{\rm L}(K) & = & -4\,g^2C_f\,\sumint_{Q}\;\Big[\kOEb \,\pOEb +\k \cdot \p \Big]\,\Delta_0(P)\, G_{\rm L}(Q)\,,\nonb\\
\bar{\Sigma}_{\rm T}(K) & = & -8\,g^2C_f\,\sumint_{Q}\;\Big[\kOEb \,\pOEb -(\hat\q \cdot \k )(\p \cdot \hat\q )\Big]\,\Delta_0(P)\,G_{\rm T}(Q)\,,
\end{eqnarray}
with $P=K-Q$. To that aim, we use the spectral representations (\ref{eq:spectral_fermion}) and (\ref{eq:spectral_gluon}). The longitudinal part gives:
\begin{eqnarray}
\bar{\Sigma}_{\rm L}(K) & = & 4\,g^2C_f\,\int_\q \int_{p_0}\;\Big[\kOEb \,p_0+\k \cdot \p \Big]\,\frac{1}{\q ^2}\,\rho_0(P)\;\frac{1}{\beta}\sum_n\frac{1}{p_0-i\omega_m}\nonb\\
& - & 4\,g^2C_f\,\int_\q \int_{p_0}\int_{q_0}\;\Big[\kOEb \,p_0+\k \cdot \p \Big]\,\rho_0(P)\, \rho_{\rm L}(Q)\;\frac{1}{\beta}\sum_n\frac{1}{(p_0-i\omega_m)(q_0-i\omega_n)}\,,\nonb\\
\end{eqnarray}
where $i\omega_m=i\omega_l-i\omega_n$ and $k_0 = i\omega_l$. The Matsubara sums give:
\begin{eqnarray}
\frac{1}{\beta}\sum_n\frac{1}{p_0-i\omega_l+i\omega_n} & = & -n_{-p_0+i\omega_l}=f_{-p_0}=\epsilon(p_0)f(|p_0|)+\mbox{vac}\,,\nonb\\
\frac{1}{\beta}\sum_n\frac{1}{(p_0-i\omega_l+i\omega_n)(q_0-i\omega_n)} & = & \frac{n_{q_0}+f_{-p_0}}{p_0+q_0-i\omega_l}=\frac{\epsilon(q_0)n(|q_0|)-\epsilon(p_0)f(|p_0|)}{p_0+q_0-i\omega_l}+\mbox{vac}\,,\nonb\\
\end{eqnarray}
where we have used $f(p_0)=\theta(-p_0)-\epsilon(p_0)f(|p_0|)$ and $n(q_0)=-\theta(-q_0)+\epsilon(q_0)n(|q_0|)$. Then, we have:
\begin{eqnarray}
\bar{\Sigma}_{\rm L,\,th}(K) & = & 4\,g^2C_f\,\int_\q \int_{p_0}\;\Big[\kOEb \,p_0+\k \cdot \p \Big]\,\frac{1}{\q ^2}\,\rho_0(P)\epsilon(p_0)f(|p_0|)\nonb\\
& - & 4\,g^2C_f\,\int_\q \int_{p_0}\int_{q_0}\;\Big[\kOEb \,p_0+\k \cdot \p \Big]\,\rho_0(P)\, \rho_{\rm L}(Q)\frac{\epsilon(q_0)n(|q_0|)-\epsilon(p_0)f(|p_0|)}{p_0+q_0-\kOE }\,,\nonb\\
\end{eqnarray}
We can now use the spectral representations backwards to write (we change variables in the second line $\q \rightarrow \p $, this is possible since the cut-off can be sent to infinity in the thermal contributions):
\begin{eqnarray}
\bar{\Sigma}_{\rm L,\,th}(K) & = & 4\,g^2C_f\,\intP \;\frac{\k \cdot \p }{\q^2}\,\sigma_0(P)\nonb\\
& - & 4\,g^2C_f\,\intP \;\Big[\kOEb \,p_0+\k \cdot \p \Big]\,\sigma_0(P)\, G_{\rm L}(K-P)\nonb\\
& + & 4\,g^2C_f\,\intQ \;\Big[\kOEb \,(\kOE -q_0)+\k \cdot \p \Big]\,\Delta_0(K-Q)\, \sigma_{\rm L}(Q)\,,
\end{eqnarray}
with $\sigma_0(P)=\epsilon(p_0)\,f_{|p_0|}\,\rho_0(P)$ and $\sigma_{\rm L}(Q)=\epsilon(q_0)\,n_{|q_0|}\,\rho_{\rm L}(Q)$. There is a static contribution which can be computed almost analytically since it does not depend on the gluon propagator:
\begin{eqnarray}
4\,g^2C_f\,\intP \;\frac{\k \cdot \p }{\q ^2}\,\sigma_0(P) & = & \frac{g^2C_f}{\pi^2}\,\int_0^\infty dp\,p\,f_p\,\int_{-1}^1 dx\,\frac{kpx}{k^2+p^2-2kpx}\nonb\\
& = & -\frac{g^2C_f}{\pi^2}\,\int_0^\infty dp\,p\,f_p\,\left[1+\frac{k^2+p^2}{4kp}\log\left(\frac{k-p}{k+p}\right)^2\right]\,.
\end{eqnarray}
In the same way, one obtains for the transverse piece:
\begin{eqnarray}
\bar{\Sigma}_{\rm T,\,th}(K) & = & -8\,g^2C_f\,\intP \;\Big[\kOEb \,p_0-(\hat\q \cdot \k )(\p \cdot \hat\q )\Big]\,\sigma_0(P)\, G_{\rm T}(K-P)\nonb\\
& + & 8\,g^2C_f\,\intQ \;\Big[\kOEb \,(\kOE -q_0)-(\hat\q \cdot \k )(\p \cdot \hat\q )\Big]\,\Delta_0(K-Q)\, \sigma_{\rm T}(Q)\,,
\end{eqnarray}
with $\sigma_{\rm T}(Q)=\epsilon(q_0)\,n_{|q_0|}\,\rho_{\rm T}(Q)$.

\section{Relation between $\Sigma_{+}$ and $\Sigma_{-}$}\label{sec:sigmarelation}

$\Sigma_{+}$ and $\Sigma_{-}$ from Eq.~(\ref{eq:sigmapm}) can be related to each other
by application of the Schwarz reflection principle: It can be shown
that $b(k_{0}=i\omega_l,k)=b^{*}(i\omega_l,k)$ is a purely real quantity
in Euclidean space (i.e.~along the imaginary axis $k_{0}=i\omega_l$),
as is $ia(i\omega_l,k)=-ia^{*}(i\omega_l,k)$. The Schwarz reflection
principle implies that in a rotation from Euclidean to Minkowski space
$k_{0}=i\omega_l\rightarrow\omega+i\epsilon$ on the one hand and $k_{0}=i\omega_l\rightarrow-\omega+i\epsilon$
on the other, the values of $b$ can be related via $b(k_{0}=\omega+i\epsilon,k)=b^{*}(-\omega+i\epsilon,k)$.
The same holds for $ia(\omega+i\epsilon,k)=-ia^{*}(-\omega+i\epsilon,k)$.
Combining the two results as in $\Sigma_{\pm}=b\pm a$ (and $\Sigma_{\pm}^{*}=b^{*}\pm a^{*}$)
gives\begin{equation}
\Sigma_{+}(\omega+i\epsilon,k)=\Sigma_{-}^{*}(-\omega+i\epsilon,k).\end{equation}
This result is of course also valid for finite $\epsilon$, i.e. all
$\omega\ge0$ and $\epsilon>0$. Particularly for $\epsilon\rightarrow0$
this implies\begin{equation}
\Re\Sigma_{+}(\omega=k,k)=\Re\Sigma_{-}(\omega=-k,k)=\frac{1}{4k}\Re\bar{\Sigma}(\omega=\pm k,k).\label{sigmacomparisons}\end{equation}


\end{document}